\documentclass[journal]{IEEEtran}
\usepackage{soul}
\usepackage{slashbox}
\usepackage{graphicx}
\usepackage{mathtools}
\usepackage{amsmath}
\usepackage{amsfonts}
\usepackage{array}
\newcommand{\PreserveBackslash}[1]{\let\temp=\\#1\let\\=\temp}
\newcolumntype{C}[1]{>{\PreserveBackslash\centering}p{#1}}
\newcolumntype{R}[1]{>{\PreserveBackslash\raggedleft}p{#1}}
\newcolumntype{L}[1]{>{\PreserveBackslash\raggedright}p{#1}}
\usepackage[usenames]{color}
\usepackage{colortbl,booktabs}
\usepackage{tabularx}
\usepackage{multicol}
\usepackage{multirow}
\usepackage{booktabs}
\usepackage[Symbol]{upgreek}
\usepackage{subfigure}
\usepackage{bm}
\usepackage{url}
\usepackage[usenames,dvipsnames,svgnames,table]{xcolor}
\usepackage{tikz}
\newcommand*\circled[1]{\tikz[baseline=(char.base)]{
            \node[shape=circle,draw,inner sep=1pt] (char) {#1};}}
\usepackage{cite}
\usepackage{balance}
\usepackage{times}

\usepackage{stfloats}
\usepackage{balance}

\usepackage[nonumberlist,acronym,shortcuts]{glossaries}
\makeglossaries
\makeindex
\makeatletter
\g@addto@macro\normalsize{%
  \setlength\abovedisplayskip{5pt}
  \setlength\belowdisplayskip{5pt}
  \setlength\abovedisplayshortskip{5pt}
  \setlength\belowdisplayshortskip{5pt}
}
\makeatother
\begin{document}


\title{\LARGE Adaptive Coding and Modulation Aided Mobile Relaying for Millimeter-Wave Flying Ad-Hoc Networks}
\author{Jiankang~Zhang,~\IEEEmembership{Senior~Member,~IEEE},
Sheng~Chen,~\IEEEmembership{Fellow,~IEEE},\\
Wei Koong Chai, ~\IEEEmembership{Senior~Member,~IEEE},
Lajos~Hanzo,~\IEEEmembership{Life Fellow,~IEEE}
\thanks{J.~ Zhang, and W.~K.~Chai are with Department of Computing \& Informatics, Bournemouth University, Poole,BH12 5BB, U.K. (E-mails: \{jzhang3, wchai\}@bournemouth.ac.uk).} %
\thanks{S.~Chen, and L.~Hanzo are with School of Electronics and Computer Science, University of Southampton, U.K.
 (E-mails: \{sqc, lh\}@ecs.soton.ac.uk).} %
\vspace*{-8mm}
}

\maketitle

\IEEEpeerreviewmaketitle

\begin{abstract}
The emerging drone swarms are capable of carrying out sophisticated
tasks in support of demanding Internet-of-Things (IoT) applications by synergistically working together. However, the target area may be out of the coverage of the ground station and it may be impractical to deploy a large number of drones in the target area due to cost, electromagnetic interference and flight-safety regulations. By exploiting the innate \emph{agility} and \emph{mobility} of unmanned aerial vehicles (UAVs), we conceive a mobile relaying-assisted drone swarm network architecture, which is capable of extending the coverage of the ground station and enhancing the effective end-to-end throughput. Explicitly, a swarm of drones forms a data-collecting drone
swarm (DCDS) designed for sensing and collecting data with the aid of their mounted cameras and/or sensors, and a powerful relay-UAV (RUAV) acts as a mobile relay for
conveying data between the DCDS and a ground station (GS). Given a time period, in order to maximize the data delivered whilst minimizing the delay imposed, we harness an $\epsilon$-multiple
objective genetic algorithm ($\epsilon$-MOGA) assisted
Pareto-optimization scheme. Our simulation results demonstrate that the
proposed mobile relaying is capable of delivering more data. As specific examples investigated in our simulations, our mobile relaying-assisted drone swarm network is capable of delivering $45.38\%$ more data than the benchmark solutions, when a stationary relay is available, and it is capable of delivering $26.86\%$ more data than the benchmark solutions when no stationary relay is available.
\end{abstract}


\section{Introduction}\label{S1}

As an emerging technology, unmanned aerial vehicles (UAVs) assisted communications have been proposed for mission-critical scenarios as well as for a range of other paradigms  \cite{zhao2019uav,Bander2020uav}. Furthermore, by autonomously forming flying ad-hoc network (FANET) \cite{arafat2022aqlearning,kim2020joint,zhang2019aeronautical} from UAVs,  the dependence on the conventional terrestrial communication infrastructure can be significantly reduced. Hence FANETs offer a promising solution both for industries and various other sectors of human life \cite{Bander2020uav} including but not limited to emergency communication \cite{zhao2019uav}, flying base station \cite{wang2019deployment} delivery, monitoring and surveillance applications in such scenarios \cite{xiao2020unmanned}.
  To elaborate, FANETs can be swiftly and flexibly deployed for providing rapid response to the above-mentioned emergency situations. Although the UAVs in an FANET are capable of communicating with each other relying on UAV-to-UAV communication links, a reliable high-rate communication solution is required to enable them to communicate with the GS, in order for them to complete their missions, including sending back the data collected by their cameras and other sensors as well as for receiving information to be disseminated. 

Typically, routing relying on multi-hop relaying is an efficient solution for exchanging information between a FANET and a GS when there are sufficiently many UAVs in the FANET for establishing at least a direct end-to-end link. Existing routing strategies may be divided into topology-based \cite{arafat2022aqlearning,tan2020research,AlKhatieb2020performance,cui2022topology} and location-based routing protocols \cite{bujari2018comparison,liu2021deep}.  Topology-based routing methods suffer either from a huge overhead required 
for maintaining a routing table or a long
delay during the route discovery process. By contrast, location-based routing protocols typically suffer from routing holes and blind path problems. Additionally, in order to establish end-to-end routing for both topology-based and location-based routing protocols, each UAV must have at least one other UAV within its communication range. Furthermore, there has to be at least one UAV which can directly communicate with the GS. In this scenario, Do \emph{et al.}~\cite{do2020UAVrelaying} investigated a UAV-based non-orthogonal multiple access (NOMA) scheme and optimized its outage by appropriately adjusting the relay-UAV's location. However, in most cases, it is challenging to deploy a large number of UAVs within a specific area, due to cost, electromagnetic interference and flight-safety regulations. When there is an insufficient number of UAVs to ensure that a direct end-to-end link's can be established or there are obstacles, such as hills or large buildings, classical stationary relaying and routing strategies will not work. The highly dynamic topology and high mobility of UAVs also impose  challenges  both on routing and on link connectivity as well as concerning the signal processing delay. As a remedy, UAVs exhibit nimble maneuverability, which makes them
eminently suitable for mobile relaying in delay-tolerant
applications. Hence, instead of relying on routing algorithms based on multi-hop relaying, we focus on the new paradigm of \emph{mobile relaying} \cite{frew2008airborne,do2021usergrouping,sharma2021outage,zeng2016throughput,zhao2020efficiency} offered by the controllable flexibility of UAVs. 

The flexibility and battery-powered nature of UAVs impose some challenges, but also offer some potential opportunities for drone-based communications and data sensing as well as data collecting. Explicitly, coverage, end-to-end throughput, power consumption and link reliability have been the key metrics to be considered, which can be maximized/minimized by optimizing the UAV's position, trajectory and charging/discharging strategy. Explicitly, Frew\emph{et al.}~\cite{frew2008airborne} proposed to load data, carry it close to destination and offload the data  with the aid of buffer on the mobile relay node, which was
termed as a `\emph{data ferry}'. Although the philosophy was pioneered by Frew, only a simple example of maintaining
a reliable communication link between a static source node (SN) and a
static destination node (DN) was provided, there is no specific network architecture design and network optimization. As an essential metric for communications, the maximization
of the throughput has attracted extensive considerations for the study of mobile relaying. Explicitly,
  Zeng \emph{et al.}~\cite{zeng2016throughput} maximized the
throughput of UAV-aided mobile relaying systems by
optimizing the source/relay transmit power along with the relay's
trajectory. As a
further development based on~\cite{zeng2016throughput}, Lin \emph{et al.}~\cite{lin2020anew}
maximized the throughput by jointly optimizing the source/relay
transmit power along with the relay's trajectory as well as the
time-slot pairing for each data packet received and
forwarded.  Li \emph{et al.}~\cite{li2021throughput} maximized the throughput in the context of UAV-assisted cognitive mobile relay networks. Explicitly, a UAV acted as a mobile relay between the primary user transmitter (PUT) and primary user receiver (PUR) as well as the secondary user transmitter (SUT) and secondary user receiver (SUR). As a further advance, the sum rate of all UAVs was maximized by Zhao \emph{et al.}~\cite{zhao2019joint} by jointly optimizing the UAV trajectory and the non-orthogonal multiple access (NOMA) precoding. As further development, Liu \emph{et al.}~\cite{liu2022throughput} maximized the average downlink throughput by jointly optimizing the UAV trajectory, the reconfigurable intelligent
surface (RIS) based passive beamforming and the source power allocation for each time slot. By contrast, 
Pang \emph{et al.}~\cite{pang2021when} proposed to deploy RIS on UAVs for maximizing the average achievable rate by jointly optimizing the trajectory and the RIS phase shifts. Additionally, mobile relaying has also been extended both to
NOMA systems~\cite{hu2022anuplink} and
to hybrid free-space optical (FSO) as well as to radio frequency (RF) systems
\cite{lee2021throughput} in order to maximize the throughput and
improve the relaying link reliability, respectively. But naturally, mobile relaying will not be sustainable if the UAV’s battery capacity is limited and no additional power supply is available. Hence, their energy efficiency was also considered by researchers. Zhao \emph{et al.}~\cite{zhao2020efficiency} aimed for maximizing the efficiency defined as weighted
sum of the energy efficiency during information transmission and the wireless power transmission efficiency. The wireless power transfer from flying energy sources to UAVs was further optimized by Oubbati \emph{et al.} \cite{oubbati2022multiagent} by relying on multiagent deep reinforcement learning.

Recently, drone swarms equipped with cameras/sensors have become a promising technology in many applications, such as video monitoring, remote sensing, disaster rescue, aerial photography and reconnaissance, which typically require high-rate communication between the drone swarms and the GS. Massive multiple-input multiple-output (MIMO) schemes relying on a large number of antennas constitute a promising solution for serving a swarm of drones in high-rate and high-reliability communications \cite{chandhar2018massive}. Explicitly, hundreds of antennas deployed at the GS are capable of focusing the energy into narrow pencil-beams for attaining huge throughput and energy efficiency improvements with the aid of transmit precoding (TPC) for the downlink (DL) and receiver combining (RC) for the uplink (UL). Research efforts have also been devoted to measure the air-to-ground channel, analyse it and remodel it by jointly considering mobility, shadowing, line-of-sight (LoS) and dynamic propagation conditions   \cite{matolak2017airground,Bithas2020uav}. The heavily-occupied sub-6 GHz frequency band becomes  not sufficient to meet ultra high-data-traffic requirements of UAV communications, the utilization of the millimeter-wave (mmWave) frequency bands has been a promising direction and feasible deployment for UAV by considering the half-wave rule of antenna theory.

\begin{table*}[!htbp]
\scriptsize
\caption{Comparison of the related mobile relaying schemes.}
\label{tab:comparison}
\begin{center} 
\begin{tabular}{|L{0.9cm}|L{1.1cm}|L{2.0cm}|L{2.5cm}|L{1.2cm}|L{1.4cm}|L{1.5cm}|L{1.7cm}|L{2.2cm}|}
\hline
       References & Optimization type & Objectives/Metrics & Optimization algorithm & Carrier Frequency & Source node & Destination node & Relaying node & Associated technology  \\ \hline
\cite{frew2008airborne} & Single-objective & Long-term throughput, time delay & Not specified & Not specified & Not specified & Not specified & Single-antenna small unmanned aircraft & Buffering \\ \hline         
\cite{do2021usergrouping} &  Single-objective & Outage probabilities Ergodic capacities & Not specified & 843 MHz & Ground users & BS & Single-antenna UAV & Energy harvesting\\ \hline
\cite{sharma2021outage} & Single-objective & Outage probabilities & Not specified & 2 GHz & Ground users & Satellite & Single-antenna UAV & Caching  \\ \hline
\cite{zeng2016throughput} & Single-objective & Throughput & Successive convex optimization & 5 GHz & Not specified & Not specified & Single-antenna UAV & Power allocation Trajectory planning  \\ \hline
\cite{zhao2020efficiency} & Single-objective & Efficiency & concave-convex procedure, penalty dual decomposition& Not specified & Single ground node & Single ground node & Single-antenna UAV & Wireless power transmission \\ \hline
\cite{li2021throughput} & Single-objective & Throughput & Lagrange dual method & Not specified &  PUT, SUT & PUR, SUR & Single-antenna UAV & Energy harvesting\\ \hline
\cite{liu2022throughput} & Single-objective & Throughput & Alternating iterative optimization & Not specified & Not specified & Not specified & RIS-UAV & Reconfigurable intelligent
surface   \\ \hline
\cite{lin2020anew} & Single-objective & Throughput & Hungary algorithm, convex approximation & 5 GHz & Ground users & BS & Single-antenna UAV & Power allocation Trajectory planning    \\ \hline
\cite{zhao2019joint} & Single-objective & Sum rate & Convex optimization & Not specified &  Ground users & BS & Single-antenna UAV & NOMA  \\ \hline
\cite{pang2021when} & Single-objective &  Average achievable rate & Successive convex approximation & Not specified &  Ground users & BS & IRS-UAV & Intelligent reflecting surface \\ \hline
\cite{hu2022anuplink} & Single-objective & Throughput & Successive convex approximation & Not specified & Ground access points & BS & Single-antenna UAV & NOMA  \\ \hline
\cite{lee2021throughput} & Single-objective & Throughput &  Successive
convex optimization & Not specified & Backhaul ground terminal & Ground user terminal & FSO/RF-aided UAV & Hybrid FSO and RF   \\ \hline
Ours & Multiple-objectives & Total data delivered, effective end-to-end throughput, time delay & $\epsilon$-MOGA & 60 GHz & Drone swarm & BS & Large-scale antenna aided UAV & mmWave and Buffering \\ \hline                                              
\end{tabular}
\end{center}
\end{table*}

However, the signals transmitted in the centimetre wave and mmWave bands can easily be blocked by obstacles \cite{rappaport2015wideband}. Thus, the communication distance becomes very limited. Additionally, most UAVs travel at a speed in the range of 30km/h to 460km/h heading in random directions, which imposes challenges in terms of their connectivity, coordination, directional communications and link adaption, etc \cite{xiao2016enabling}. Furthermore, it is challenging to implement the traditional link adaptation of adaptive coding and modulation (ACM), which relies on the near-instantaneous signal-to-noise-ratio (SNR) in practical aeronautical communications. This is because it is required to frequently estimate the instantaneous SNR and frequently change the ACM mode. Hence, advanced TPC and link adaptation schemes have to be conceived for tackling these challenges of drone swarm communications.  Our distance-based ACM of \cite{zhang2018adaptive} is capable of supporting high-rate aeronautical communication with a quickly judgement threshold of communication distance.

In Table~\ref{tab:comparison} we boldly and explicitly compare the main
contributions of \cite{frew2008airborne,do2021usergrouping,sharma2021outage,zeng2016throughput,
zhao2020efficiency,li2021throughput,liu2022throughput,lin2020anew,zhao2019joint,
pang2021when,hu2022anuplink,lee2021throughput}, which have the most similar objective of relaying data from a source node to a destination node. By observing Table~I, we can see that existing mobile relaying solutions harness a UAV for mobile relaying between a GS and the terminal users/sensors either to maximize the end-to-end throughput or to reduce the outage probability, as comparative studied in Table~\ref{tab:comparison}. But as a prerequisite, the UAV relay should be able to establish a direct communication link both with the GS and the terminal users/sensors. When the terminal users/sensors are far away from the GS and hence it is impossible to build up a communication link harnessing a single relay node, multiple relay nodes have to be deployed, which will impose a great challenge on the network in terms of cost, network design, configuration, and network optimization. The 'data ferry' philosophy pioneered by Frew~\cite{frew2008airborne} is capable of periodically ferrying data from the source node to the destination node even if the relay node cannot maintain direct communication with the source node and destination node at the same time. But again, there is no comprehensive network design, aiming for network optimization. Inspired by the \emph{data ferry} philosophy in the open literature, we conceive a distance-based ACM scheme for relay-assisted drone swarm communications relying on mmWave massive MIMO solutions, which harness a swarm of small or micro drones for data sensing/collecting of that relies on a powerful fixed wing UAV as a mobile relaying node. The powerful fixed wing UAV circulates between the data sensing/collecting drones and the GS to relay data to the GS. Furthermore, we jointly optimize multiple objectives with the aid of our Pareto-optimization algorithm, rather than optimizing a single objective or converting multiple objectives into alternative sub-optimization problems.  
 our contributions are summarized as follows:
\begin{itemize}
\item [1)] We propose a UAV communication architecture consisting of a data-collecting drone swarm (DCDS), a relay-UAV (RUAV) and a GS. Explicitly, the DCDS collect information via their cameras or sensors across the target-area, whilst the RUAV, equipped with a large-scale mmWave antenna, acts as a mobile relay for conveying data between the DCDS and GS.
\item[2)] We conceive a distance-based ACM and relay-assisted drone swarm communication scheme by switching the ACM modes based on the communication distance and exploiting the controllable-mobility of the RUAV. Explicitly, the channel qualities between RUAV and GS as well as between RUAV and DCDS are dominated by the communication distance, since the channel exhibits Rician fading instead of Rayleigh fading due to the high altitude of RUAVs.
\item [3)] We propose a buffer-aided mobile-relay-assisted drone swarm communication protocol for the challenging scenario, where the DCDS-to-RUAV and RUAV-to-GS links do not exist concurrently. We define the \emph{effective end-to-end throughput} metric, which is then used as one of the multiple objectives of the optimization problem formulated. Furthermore, in order to  maximize the effective end-to-end throughput and to simultaneously minimize the delay  imposed, we develop an  $\epsilon$-multiple objective genetic algorithm ($\epsilon$-MOGA) assisted Pareto-optimization scheme for jointly optimizing the data uploading and offloading points, the maximum factor of caching data, and the minimum factor of offloading data, given a specific buffer size as well as distance between the GS and DCDS.\footnote{The factors of caching and off-loading controlling the DCDS and RUAV actions will be explicitly exemplified later.}
\end{itemize}
 
The rest of this paper is organized as follows. Section~\ref{S2} presents on mobile relaying aided drone swarm mmWave communications model. Both the throughput of the DCDS-to-RUAV link and the throughput of the RUAV-to-GS link are analyzed in Section~\ref{S3}. In Section~\ref{S4}, the multiple-objective optimization problem of relaying assisted FANETs is formulated, which includes both static relaying and mobile relaying scenarios. Our $\epsilon$-MOGA assisted Pareto-optimization scheme is also developed in this section. The implementation issue and computational complexity are also discussed. Section~\ref{S5} is devoted to simulation experiments, which includes the scenarios when a the stationary relay is available and when it is not available. In Section~\ref{S6}, we conclude and briefly discuss our future research ideas.

\begin{figure*}[bp!]
\vspace*{-2mm}
\begin{center}
 \includegraphics[width=0.65\textwidth]{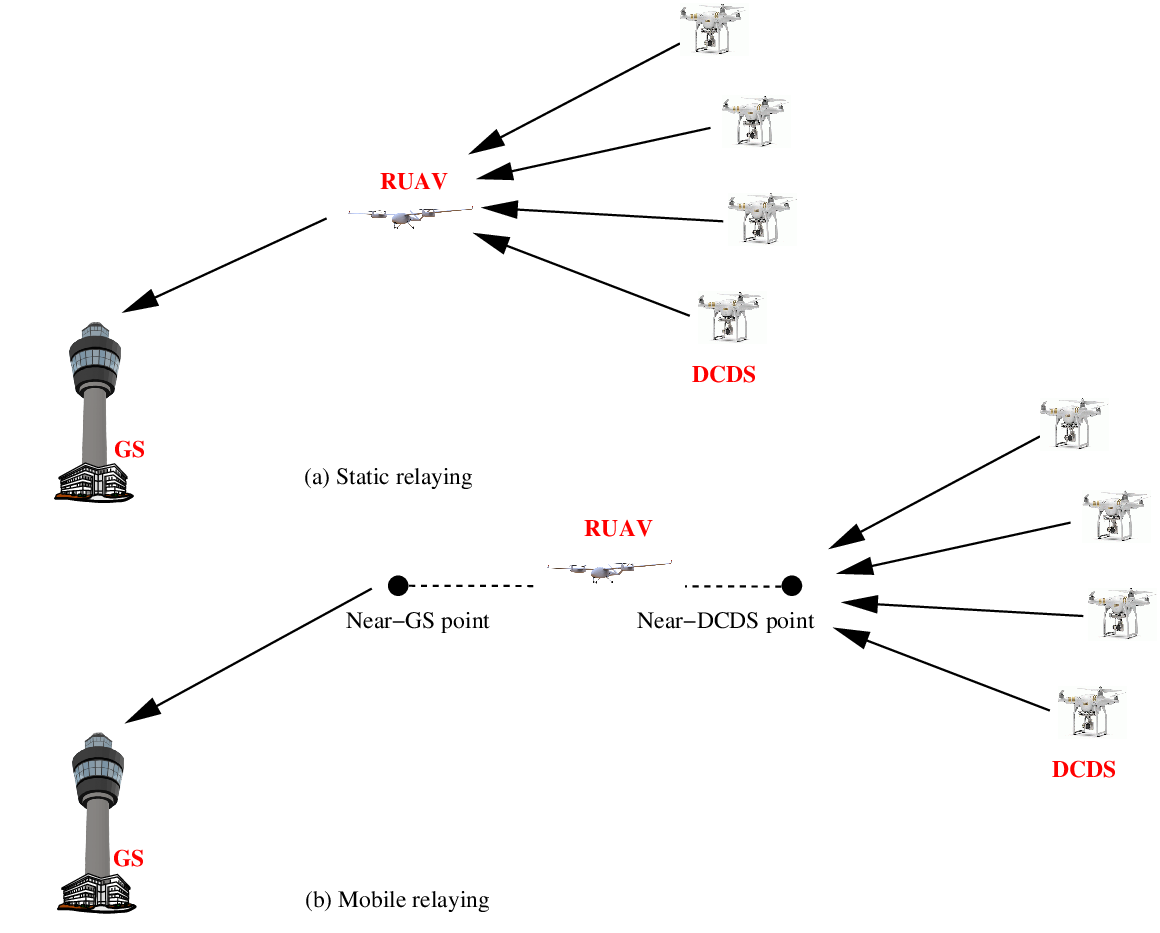}
\end{center}
\vspace*{-5mm}
\caption{UAV-relay aided drone swarm communications.}
\label{FIG1}
\vspace*{-1mm}
\end{figure*}

\section{System model}\label{S2}

We assume that a drone swarm is supported by a UAV-relay, where a GS centrally processes the signals collected by the remote drone swarms, and each drone swarm is served by the relay UAV (RUAV). We also assume that the GS is capable of simultaneously serving $B$ RUAVs and the  corresponding $B$ drone swarms by relying on the ubiquitous orthogonal frequency-division multiplexing access (OFDMA) protocol for supporting the $B$ RUAVs and drone swarms.

Furthermore, a RUAV is simultaneously serving $K$ drones, that is, there are $K$ drones in a drone swarm. More specifically, we illustrate an end-to-end link between a drone swarm and the GS in Fig.~\ref{FIG1}, where the drone swarms collect data via their cameras or other sensors in the target area, whilst a RUAV actively relays the collected data to the GS for central signal processing. Explicitly, each drone only has a single antenna due to its limited constrained fuselage size and its affordable energy. A drone swarm consisting of $K$ drones simultaneously transmits its collected information to a RUAV, which is more powerful in terms of its flying speed, energy and communication equipment. Furthermore, the RUAV also utilizes the OFDMA protocol for receiving and forwarding the DCDSs' data, which allows it to simultaneously transmit and receive data at the same time without jamming its own received signal by its own transmitted signals. The RUAV has $N_{\text{total}}$ antennas, of which $N_{r}$ antennas are data-receiving antennas (DRAs) utilized for receiving data, whilst $N_{t}$ antennas are data-transmitting antennas (DTAs) used for sending data. We assume that $N_{t} = K \le N_{r} \le N_{\text{total}}$, in line with the maximum attainable spatial degrees of freedom for relaying the DCDS's data. The GS has a large-scale antenna array, having $N_{g}$ DTAs. The length of the cyclic prefix (CP) $N_{cp}$ is higher than the channel impulse response (CIR), which indicates that there is no inter-symbol interference and the receiver can process the signals on a subcarrier-by-subcarrier basis. To simplify notations, we will omit the OFDM symbol index and the subcarrier index in our investigation. The end-to-end communication link between the DCDS and GS consists of the DCDS-to-RUAV link and the RUAV-to-GS link, which will be elaborated on in Subsection~\ref{S2-1} and Subsection~\ref{S2-2}, respectively.

\subsection{Signal model of DCDS-to-RUAV}\label{S2-1}

The discrete signals received at the RUAV $\bm{r} \in \mathbb{C}^{N_{r}}$ can be formulated as
\begin{align}\label{eq1}
\bm{r} =& \bm{H}_{0}^{(d/r)}\big(\bm{P}_{{\rm rx},0}^{(d,x)}\big)^{\frac{1}{2}}
  \bm{s}_{0} + \sum\limits_{a=1}^A  \bm{H}_{a}^{(d/r)}
  \big(\bm{P}_{{\rm rx},a}^{(d,x)}\big)^{\frac{1}{2}} \bm{s}_{a} +\bm{n} ,
\end{align}
where $\bm{H}_a^{(d/r)}\! \in\! \mathbb{C}^{N_r\times K}$ represents the uplink MIMO channel between the $a^{\text{th}}$ drone swarm and the RUAV, $\bm{s}_a\! =\! \big[s^{(a)}_1 ~ s^{(a)}_2 \cdots s^{(a)}_K\big]^{\rm T}$ is the $a^{\text{th}}$ drone swarm's transmit signal vector having a normalized transmit power of $\mathcal{E}\left\{\bm{s}_{a}\bm{s}_{a}^{\rm H}\right\}\! =\! \bm{I}_K$, whilst $\bm{n}\! \in\! \mathbb{C}^{N_{r}}$ is the additive white Gaussian noise (AWGN) with zero mean vector and covariance matrix of $\sigma_{n}^{2}\bm{I}_{N_{r}}$, and $\bm{P}_{{\rm rx},a}^{(d,x)}\! =\! \text{diag}\Big\{P_{{\rm rx},a(1)}^{(d,x)}, \cdots , P_{{\rm rx},a(K)}^{(d,x)}\Big\}$ are the powers of the $K$ drones' signals received at the RUAV. In (\ref{eq1}), the subscript $a = 0$ denotes the desired drone swarm, whilst the subscript $a\in\{1,2,\cdots, A\} $ denotes the $a^{\text{th}}$ co-channel drone swarm contaminating the desired one, with $A$ being the number of interfering drone swarms. Clearly, $A \le B-1$. Furthermore, in the superscript,  $x = p$ represents pilot training and $x = s$ denotes data transmission. 

Still referring to (\ref{eq1}), the received power $P_{{\rm rx},a(k)}^{(d,x)}$ is a function of the transmit power and path loss, which is given by
\begin{align}\label{eq2}
P_{{\rm rx},a(k)}^{(d,x)} =& P_{{\rm tx},a(k)}^{(d,x)}10^{-0.1L_{\text{path loss}, a,k}} ,
\end{align}
where the path loss model $L_{\text{path loss}, a}$ of mmWave signals is modelled as \cite{rappaport2015wideband}
\begin{align}\label{eq3}
L_{\text{path loss}, a} [\text{dB}] =&  \alpha + \beta 10 \log_{10}\left(d_{a,k}\right) + L_{\sigma}.
\end{align}
In (\ref{eq3}), $\alpha$ is the is the path loss in decibels (dB) at the reference distance $d_{0}$ calculated using the Friis free-space path loss model, $\beta$ is the linear slope, $d_{a,k}$ is the distance between the RUAV and the $k$th drone of the $a^{\text{th}}$ drone swarm in meter, and $L_{\sigma}$ is the shadow fading \cite{rappaport2002wireless}, which is a zero-mean Gaussian random variable with standard deviation $\sigma$ in dB.

\begin{figure*}[tp!]\setcounter{equation}{6}
\vspace*{-4mm}
\begin{align}\label{eq7}
\bm{vec}\big(\widehat{\bm{H}}_{a}^{(d/r)}\big) =& \nu_{r} \bm{vec}\Big(\bm{H}_{a, {\rm d}}^{(d/r)}\Big)\! +\! \zeta_{r}^{2} \bm{\Psi}_{\bm{H}_{a, {\rm r}}^{(d/r)}}\bm{\Phi}_{\bm{H}_{a, {\rm r}}^{(d/r)}} \! 
  \bigg(\zeta_{r} \bm{vec}\Big(\bm{H}_{a, {\rm r}}^{(d/r)}\Big) \nonumber \\
& + \zeta_{r}\!\!\!\! \sum\limits_{a' = 0, a' \neq a}^{A}\!\!\! \bm{vec}\Big(\! \bm{H}_{a', {\rm r}}^{(d/r)}
  \big(\bm{P}_{{\rm rx},a'}^{(d,p)}\big)^{\frac{1}{2}}\big(\bm{P}_{{\rm rx},a}^{(d,p)}\big)^{-\frac{1}{2}}\Big)\! +\! \bm{vec}\Big(\! \bar{\bm{N}}\big(\bar{\bm{S}}^{(p)}\big)^{\rm H}\big(\bm{P}_{{\rm rx},a}^{(d,p)}\big)^{-\frac{1}{2}}\Big)\!\! \bigg) ,\!
\end{align}
\hrulefill
\vspace*{-4mm}
\end{figure*}

The DCDS-to-RUAV MIMO channel is an air-to-air channel, which is dominated by its line-of-sight (LoS) component, but scattered components may still exist that impinge from the reflections  mountains/buildings etc. Hence, the DCDS-to-RUAV MIMO channel is modeled as a Rican channel, which is formulated as
\setcounter{equation}{3}
\begin{align}\label{eq4}
\bm{H}_{a}^{(d/r)} =& \nu_{r} \bm{H}_{a, {\rm d}}^{(d/r)} + \zeta_{r} \bm{H}_{a, {\rm r}}^{(d/r)} ,
\end{align}
where we have $\nu_{r} = \sqrt{K_{\text{Rice},r}/ \left(1 + K_{\text{Rice},r}\right)}$ and $\zeta_{r} = 1 / \left(1 + K_{\text{Rice},r}\right)$ with $K_{\text{Rice},r}$ being the Rician factor, while $\bm{H}_{a, {\rm d}}^{(d/r)}$ is the deterministic part of the Rician channel and $\bm{H}_{a, {\rm r}}^{(d/r)}$ is the scattered component of the Rician channel. Typically, the Rician factor is affected by the altitude of the UAV \cite{zhang2019aeronautical,matolak2017airground,haas2002aeronautical}, where a higher UAV is more likely experience a higher Rician factor, namely a stronger LoS component. Owing to the minimum safety separation distance, the transmit antennas on different drones experience  uncorrelated fading, whilst the receive antennas deployed on URAVs are located at the same site.  Hence there may exist correlation between the $N_r$ DRAs. Therefore, the scattered component $\bm{H}_{a, {\rm r}}^{(d/r)}$ can be formulated as
\begin{align}\label{eq5}
\bm{H}_{a, {\rm r}}^{(d/r)} =& \left(\bm{R}_{{\rm rx}, a}^{d/r}\right)^{\frac{1}{2}}\bm{G}_{a}^{d/r} , 
\end{align}
where $\bm{R}_{{\rm rx}, a}^{d/r}\! \in\! \mathbb{C}^{N_{r} \times N_{r}}$ is the spatial correlation matrix of the $N_{r}$ DRAs, and the entries of $\bm{G}_{a}^{d/r}\! \in\! \mathbb{C}^{N_{r} \times K}$ are independent and identically distributed (i.i.d.) complex random variables obeying the distribution $\mathcal{CN}(0,1)$.

The transmitted vector $\bm{s}_{0}$ can be detected by computing the inner product between the received vector $\bm{r}$ and a linear receiver combing (RC) matrix $\bm{W}^{(d/r)}_0$, which is expressed as 
\begin{align}\label{eq6}
\widehat{\bm{s}}_0 =& \sqrt{\lambda^{(d/r)}} \bm{W}_0^{(d/r)} \bm{H}_0^{(d/r)} \big(\bm{P}_{{\rm rx},0}^{(d,x)}\big)^{\frac{1}{2}} \bm{s}_0 \nonumber \\
&+ \sqrt{\lambda^{(d/r)}} \bm{W}_0^{(d/r)} \sum\limits_{a=1}^A \bm{H}_a^{(d/r)}
\big(\bm{P}_{{\rm rx},a}^{(d,x)}\big)^{\frac{1}{2}} \bm{s}_a + \widetilde{\bm{n}} ,
\end{align}
where $\lambda^{(d/r)}\!=\!\frac{1}{K}\text{Tr}\Big\{\mathcal{E}\Big\{\bm{W}_{0}^{(d/r)} \big(\bm{W}_{0}^{(d/r)}\big)^{\rm H}\Big\}\Big\}$ is a normalization factor, and $\widetilde{\bm{n}}\! =\! \sqrt{\lambda^{(d/r)}}\bm{W}_{0}^{(d/r)}\bm{n}$ is the effective noise after applying the RC operation. The RC matrix based on the classical matched filter (MF) is given by $\bm{W}_0^{(d/r)}\! =\! \left(\widehat{\bm{H}}_0^{(d/r)}\right)^{\rm H}$, where $\widehat{\bm{H}}_0^{(d/r)}$ is the estimate of $\bm{H}_0^{(d/r)}$. Upon using the optimal minimum mean square error (MMSE) channel estimator \cite{kay2003fundamentals}, the channel estimate $\widehat{\bm{H}}_a^{(d/r)}$, $a = 0, 1, \cdots, A$, is given by Eq.~(\ref{eq7}), 
where $\bar{\bm{S}}^{(p)}\!\in \! \mathbb{C}^{K \times K}$ is the pilot symbol matrix associated with $\bar{\bm{S}}^{(p)}\big(\bar{\bm{S}}^{(p)}\big)^{\rm H}\! =\! \bm{I}_{K}$, and $\bar{\bm{N}}\! \in\! \mathbb{C}^{N_{r} \times K}$ is the noise matrix over $K$ pilots, while $\bm{\Psi}_{\bm{H}_{a, {\rm r}}^{(d/r)}}$ is the covariance matrix of $\bm{vec}\big(\bm{H}_{a, {\rm r}}^{(d/r)}\big)$ given by
\setcounter{equation}{7}
\begin{align}\label{eq8}
\bm{\Psi}_{\bm{H}_{a, {\rm r}}^{(d/r)}}\!\! =& \mathcal{E}\left\{\bm{vec}\left(\bm{H}_{a, {\rm r}}^{(d/r)}\right) \bm{vec}\left(\bm{H}_{a, {\rm r}}^{(d/r)}\right)^{ \rm H}\right\} \!\!=\!\! \bm{I}_{K} \otimes \bm{R}_{{\rm rx}, a}^{d/r} ,
\end{align}
and $\bm{\Phi}_{\bm{H}_{a, {\rm r}}^{(d/r)}}$ in (\ref{eq7}) is defined by
\begin{align}\label{eq9}
\bm{\Phi}_{\bm{H}_{a, {\rm r}}^{(d/r)}} =& \bigg(\sigma_{n}^2 \big(\bm{P}_{{\rm rx}, a}^{(r,p)}\big)^{-1}
  \! \otimes\! \bm{I}_{N_{r}}\! +\! \zeta_{r}^{2}\bm{\Psi}_{\bm{H}_{a, {\rm r}}^{(d/r)}}	\nonumber \\
  & +  \zeta_{r}^{2}\sum\limits_{a' = 0,a'\neq a}^{A}\!\widetilde{\bm{P}}_{{\rm rx}, a'}^{(r,p)}\bm{\Psi}_{\bm{H}_{a, {\rm r}}^{(d/r)}}\bigg)^{-1} ,
\end{align}
in which $\widetilde{\bm{P}}_{{\rm rx}, a'}^{(r,p)}$ is given as
\begin{align}\label{eq10}
\widetilde{\bm{P}}_{{\rm rx}, a'}^{(r,p)} = \left(\bm{P}_{{\rm rx}, a'}^{(r,p)}\big(\bm{P}_{{\rm rx}, a}^{(r,p)}\big)^{-1}\right) \otimes \bm{I}_{K} .
\end{align}

 \begin{figure*}[tp!]\setcounter{equation}{17}
\vspace*{-4mm}
\begin{align}\label{eq18}
\bm{vec}\big(\widehat{\bm{H}}_{b}^{(r/g)}\big)\! =& \nu_{g} \bm{vec}\big(\bm{H}_{b,{\rm d}}^{(r/g)}\big) + \zeta_g^2 \bm{\Psi}_{\bm{H}_{b,{\rm r}}^{(r/g)}}\bm{\Phi}_{\bm{H}_{b, {\rm r}}^{(r/g)}} \bigg(\zeta_{g} \bm{vec}\big(\bm{H}_{b, {\rm r}}^{(r/g)}\big) \nonumber \\
& + \zeta_g\! \sqrt{\frac{P_{{\rm rx},b'}^{(r,p)}}{P_{{\rm rx},b}^{(r,p)}}} \sum\limits_{b' = 0, b' \neq b}^{B-1}\!\! \bm{vec}\big(\bm{H}_{b', {\rm r}}^{(r/g)} \big) + \sqrt{\frac{1}{P_{{\rm rx},b}^{(r,p)}}}\bm{vec}\big(\bar{\bm{V}}(\bar{\bm{X}}^{(p)})^{\rm H}\big)\bigg) , \!
\end{align}
\hrulefill
\vspace*{-4mm}
\end{figure*}

The true channel $\bm{vec}\big(\bm{H}_{a}^{(d/r)}\big)$ is equal to the MMSE estimate $\bm{vec}\big(\widehat{\bm{H}}_{a}^{(d/r)}\big)$ plus the channel estimation error $\bm{vec}\big(\widetilde{\bm{H}}_{a}^{(d/r)}\big)$:
\setcounter{equation}{10}
\begin{align}\label{eq11}
\bm{vec}\big(\bm{H}_{a}^{(d/r)}\big) = & \bm{vec}\big(\widehat{\bm{H}}_{a}^{(d/r)}\big) + \bm{vec}\big(\widetilde{\bm{H}}_{a}^{(d/r)}\big) .
\end{align}
Clearly, $\bm{vec}\big(\widetilde{\bm{H}}_{a}^{(d/r)}\big)$ is independent of both $\bm{vec}\big(\bm{H}_{a}^{(d/r)}\big)$ and $\bm{vec}\big(\widehat{\bm{H}}_{a}^{(d/r)}\big)$, and it obeys the distribution $\mathcal{CN}\left(\bm{0}_{N_{r}K}, \bm{\Psi}_{\widetilde{\bm{H}}_{a, {\rm r}}^{(d/r)}}\right)$ with the covariance matrix $\bm{\Psi}_{\widetilde{\bm{H}}_{a, {\rm r}}^{(d/r)}}$ given by
\begin{align}\label{eq12}
\bm{\Psi}_{\widetilde{\bm{H}}_{a, {\rm r}}^{(d/r)}} =& \zeta_{r}^{2}\bm{\Psi}_{\bm{H}_{a, {\rm r}}^{(d/r)}} - \zeta_{r}^{2}\bm{\Psi}_{\widehat{\bm{H}}_{a, {\rm r}}^{(d/r)}} .
\end{align}
The covariance matrix of the MMSE channel estimate $\bm{\Psi}_{\widehat{\bm{H}}_{a, {\rm r}}^{(d/r)}}$ in (\ref{eq12}) is given by
\begin{align}\label{eq13}
\bm{\Psi}_{\widehat{\bm{H}}_{a, {\rm r}}^{(d/r)}} =& \bm{\Psi}_{\bm{H}_{a, {\rm r}}^{(d/r)}}\bm{\Phi}_{\bm{H}_{a, {\rm r}}^{(d/r)}}\bm{\Psi}_{\bm{H}_{a, {\rm r}}^{(d/r)}} .
\end{align}

\subsection{Signal model of RUAV-to-GS}\label{S2-2}

The $K$ drones' signals are detected and forwarded by the RUAV to the GS with aid of its $K$ DTAs. Since there are $B$ RUAVs, the signal vector received at the GS can be written as
\begin{align}\label{eq14}
\bm{y} = & \sqrt{P_{{\rm rx},0}^{(r,x)}}\bm{H}_{0}^{(r/g)} \bm{x}_{0} + \sqrt{P_{{\rm rx},b}^{(r,x)}}\sum\limits_{b=1}^{B-1}  \bm{H}_{b}^{(r/g)} \bm{x}_{b} +\bm{v} ,
\end{align}
where $b\! =\! 0$ indicates the desired RUAV, and $b\! \neq\! 0$ refer to the  interfering RUAVs, while $\bm{H}_{b}^{(r/g)}\! \in\! \mathbb{C}^{N_{g} \times K}$ is the MIMO channel matrix between the $b$th RUAV and the GS, $\bm{x}_b\! \in\! \mathbb{C}^K$ is the $b$th transmitted signal vector, and $\bm{v}$ is the zero-mean AWGN vector having a covariance matrix of $\sigma_v^2\bm{I}_{N_g}$. Because the DTAs of the each RUAV are co-located, at the GS, the received powers of the $K$ DTAs are almost the same, and approximately $P_{{\rm rx},b(1)}^{(r,x)}\! =\! P_{{\rm rx},b(2)}^{(r,x)}\! =\! \cdots\! =\! P_{{\rm rx},b(K)}^{(r,x)}\! =\! P_{{\rm rx},b}^{(r,x)}$. Again, the subscript $x\! =\! p$ denotes pilot training and $x\! =\! s$ represents data transmission. The value of $P_{{\rm rx},b}^{(r,x)}$ can be acquired following the path loss model of (\ref{eq2}) and (\ref{eq3}) associated with a different shadowing factor $L_{\sigma}^{(g)}$, since the local environment of the GS is different from that of the RUAV.

The air-to-ground channel of the RUAV-to-GS link is also Rician, but it suffers from stronger scattering and reflection. Hence, the MIMO channel matrix $\bm{H}_{b}^{(r/g)}$ can be expressed as
\begin{align}\label{eq15}
\bm{H}_{b}^{(r/g)} =& \nu_{g} \bm{H}_{b, {\rm d}}^{(r/g)} + \zeta_{g} \bm{H}_{b, {\rm r}}^{(r/g)} ,
\end{align}
where $\nu_{g} = \sqrt{K_{\text{Rice},g}/ \left(1 + K_{\text{Rice},g}\right)}$ and $\zeta_{g} = 1 / \left(1 + K_{\text{Rice},g}\right)$ with $K_{\text{Rice},g}$ being the Rician factor, while $\bm{H}_{b, {\rm d}}^{(r/g)}$ is the deterministic part of the Rician channel and $\bm{H}_{b, {\rm r}}^{(r/g)}$ is the scattered component of the Rician channel. Since the RUAV has $N_{\text{total}}$ antennas, which is much more than $K$, it can always select $K$ uncorrelated DTAs for forwarding its drone swarm's signals to the GS. Again, the scattered component $\bm{H}_{b, {\rm r}}^{(r/g)}$ can be expressed as
\begin{align}\label{eq16}
\bm{H}_{b, {\rm r}}^{(r/g)} =& \left(\bm{R}_{{\rm rx}, b}^{r/g}\right)^{\frac{1}{2}}\bm{G}_{b}^{r/g} ,
\end{align}
where $\bm{R}_{{\rm rx}, b}^{r/g}\! \in\! \mathbb{C}^{N_{g} \times N_{g}}$ is the spatial correlation matrix of the $N_{g}$ DRAs and $\bm{G}_{b}^{r/g}\! \in\! \mathbb{C}^{N_{g} \times K}$ has i.i.d. complex entries obeying the distribution $\mathcal{CN}(0,1)$.

Similar to the DCDS-to-RUAV signal model, the estimate of $\bm{x}_{0}$ can be acquired by applying the MF-based RC, which is given by
\begin{align}\label{eq17}
\bm{x}_{0} =& \sqrt{\lambda^{(r/g)}P_{{\rm rx},0}^{(r,x)}}\bm{W}_{0}^{(r/g)}\bm{H}_{0}^{(r/g)}
  \bm{x}_{0} \nonumber \\
  & + \sqrt{\lambda^{(r/g)}}\bm{W}_{0}^{(r/g)}\sum\limits_{b=1}^{B-1} \sqrt{P_{{\rm rx},b}^{(r,x)}} \bm{H}_{b}^{(r/g)}  \bm{x}_{b} + \widetilde{\bm{v}} ,
\end{align}
where $\lambda^{(r/g)}\!=\!\frac{1}{K}\text{Tr} \!\Big\{ \!\mathcal{E} \!\Big\{\bm{W}_{0}^{(r/g)} \!
 \big(\bm{W}_{0}^{(r/g)}\big)^{\rm H} \!\Big\} \!\Big\}$ is the normalization factor, and $\widetilde{\bm{v}} \!=\! \sqrt{\lambda^{(r/g)}}\bm{W}_{0}^{(r/g)}\bm{v}$ is the effective noise. The MF-based RC matrix is given by $\bm{W}_{0}^{(r/g)} = \left(\widehat{\bm{H}}_{0}^{(r/g)}\right)^{\rm H}$, and the MMSE channel estimate $\widehat{\bm{H}}_{b}^{(r/g)}$ of the true channel $\bm{H}_{b}^{(r/g)}$ is given by Eq.~(\ref{eq18}), 
where $\bar{\bm{X}}^{(p)}\! \in\! \mathbb{C}^{K\times K}$ is the pilot symbol matrix associated with $\bar{\bm{X}}^{(p)}(\bar{\bm{X}}^{(p)})^{\rm H}\! =\! \bm{I}_K$, and $\bar{\bm{V}}\! \in\! \mathbb{C}^{N_g\times K}$ is the noise matrix over $K$ pilots, while the covariance matrix $\bm{\Psi}_{\bm{H}_{b, {\rm r}}^{(r/g)}}$ of $\bm{vec}\big(\bm{H}_{b}^{(r/g)}\big)$ is given by
\setcounter{equation}{18}
\begin{align}\label{eq19}
\bm{\Psi}_{\bm{H}_{b, {\rm r}}^{(r/g)}} \!\!=\!\! \mathcal{E}\left\{\bm{vec}\left(\bm{H}_{b, {\rm r}}^{(r/g)}\right) \bm{vec}\left(\bm{H}_{b, {\rm r}}^{(r/g)}\right)^{ \rm H}\right\}\!\! =\!\! \bm{I}_{K} \otimes \bm{R}_{{\rm rx}, b}^{r/g} ,
\end{align}
and $\bm{\Phi}_{\bm{H}_{b, {\rm r}}^{(r/g)}}$ in (\ref{eq18}) is formulated as 
\begin{align}\label{eq20}
\bm{\Phi}_{\bm{H}_{b, {\rm r}}^{(r/g)}} =& \bigg(\frac{\sigma_v^2}{P_{{\rm rx}, b}^{(r,p)}}
  \! \otimes\! \bm{I}_{N_g K}\! +\! \zeta_{g}^{2}\bm{\Psi}_{\bm{H}_{b, {\rm r}}^{(r/g)}} \nonumber\\
  &	+  \frac{\zeta_{g}^{2}}{P_{{\rm rx}, b}^{(r,p)}}\sum\limits_{b' = 0,b'\neq b}^{B-1} P_{{\rm rx}, b'}^{(r,p)}\bm{\Psi}_{\bm{H}_{b', {\rm r}}^{(r/g)}}\bigg)^{-1} .
\end{align}
More specifically, the true channel $\bm{vec}\big(\bm{H}_{b}^{(r/g)}\big)$ is  given by
\begin{align}\label{eq21}
\bm{vec}\big(\bm{H}_{b}^{(r/g)}\big) = & \bm{vec}\big(\widehat{\bm{H}}_{b}^{(r/g)}\big) + \bm{vec}\big(\widetilde{\bm{H}}_{b}^{(r/g)}\big) ,
\end{align}
and the channel estimation error obeys $\bm{vec}\big(\widetilde{\bm{H}}_{b}^{(r/g)}\big)\! \sim\! \mathcal{CN}\left(\bm{0}_{N_{g}K}, \bm{\Psi}_{\widetilde{\bm{H}}_{b, {\rm r}}^{(r/g)}}\right)$, with $\bm{\Psi}_{\widetilde{\bm{H}}_{b, {\rm r}}^{(r/g)}}$ given by
\begin{align}\label{eq22}
\bm{\Psi}_{\widetilde{\bm{H}}_{b, {\rm r}}^{(r/g)}} =& \zeta_{g}^{2}\bm{\Psi}_{\bm{H}_{b, {\rm r}}^{(r/g)}} - \zeta_{g}^{2}\bm{\Psi}_{\widehat{\bm{H}}_{b, {\rm r}}^{(r/g)}} , 
\end{align}
and
\begin{align}\label{eq23}
\bm{\Psi}_{\widehat{\bm{H}}_{b, {\rm r}}^{(r/g)}} =& \bm{\Psi}_{\bm{H}_{b, {\rm r}}^{(r/g)}}\bm{\Phi}_{\bm{H}_{b, {\rm r}}^{(r/g)}}\bm{\Psi}_{\bm{H}_{b, {\rm r}}^{(r/g)}} .
\end{align}

\section{Analysis of the achievable throughput}\label{S3}

Here, we use the decode and forward relaying protocol as an example for analyzing the achievable uplink throughput. Clearly, the end-to-end uplink throughput is the minimum of the DCDS-to-RUAV link's throughput and the RUAV-to-GS link's throughput. 

\subsection{The achievable throughput of the DCDS-to-RUAV link}\label{S3.1}

The ergodic achievable uplink throughput of the $k$th drone for the targeted DCDS is formulated as
\begin{align}\label{eq24}
C_{k}^{(d/r)} = \mathcal{E}\left\{\log_{2}\left(1 + \frac{P_{\text{S}, k}^{(d/r)}}{P_{\text{IN}, k}^{(d/r)}}\right)\right\} ,
\end{align}
where $P_{\text{S}, k}^{(d/r)}$ and $P_{\text{IN}, k}^{(d/r)}$ are the powers of the desired signal and of the interference-plus-noise, respectively. By invoking \emph{Lemma~I} of \cite{zhang2014power}, $C_{k}^{(d/r)}$ in (\ref{eq18}) can be approximated as 
\begin{align}\label{eq25}
C_{k}^{(d/r)} \approx \log_{2}\left(1 + \frac{\bar{P}_{\text{S}, k}^{(d/r)}}{\bar{P}_{\text{IN}, k}^{(d/r)}}\right) ,
\end{align}
where $\bar{P}_{{\rm S},k}^{(d/r)}\! =\! \mathcal{E}\left\{P_{{\rm S},k}^{(d/r)}\right\}$ and $\bar{P}_{{\rm IN},k}^{(d/r)}\! =\! \mathcal{E}\left\{P_{{\rm IN},k}^{(d/r)}\right\}$. 

Let $k^{*}$ represent the investigated drone of the desired drone swarm upon invoking \emph{Lemma~1} of \cite{fernandes2013inter} and \emph{Lemma~2} of \cite{hoydis2012random}, $\mathcal{E}\left\{P_{{\rm S},k^{*}}^{(d/r)}\right\}$ can be derived as
\begin{align}\label{eq26}
\mathcal{E}\!\!\left\{P_{{\rm S},k^{*}}^{(d/r)}\!\!\right\} \!\!=& \!\lambda^{(d/r)}P_{{\rm rx},0(k^{*})}^{(d,s)}\! \bigg(\! \text{Tr}\bigg\{\! \nu_{r}^{2}\bm{B}_{\bm{H}_{0, {\rm d}, [~:k^{*}]}^{(d/r)}}\!\!  + \zeta_{r}^{2}\bm{\Psi}_{\widehat{\bm{H}}_{0, {\rm r}, [~:k^{*}]}^{(d/r)}}\!\! \nonumber \\
&+ \zeta_{r}^{2} \bm{\Psi}_{\widehat{\bm{H}}_{0,{\rm r},[~:k^{*}]}^{(d/r)}} \bm{\Psi}_{\widetilde{\bm{H}}_{0,{\rm r},[~:k^{*}]}^{(d/r)}} \bigg\}\! \bigg)^{2}\! ,\!
\end{align}
where $\bm{B}_{\bm{H}_{0,{\rm d},[~:k^{*}]}^{(d/r)}}\! =\! \bm{H}_{0,{\rm d},[~:k^{*}]}^{(d/r)}\big(\bm{H}_{0,{\rm d},[~:k^{*}]}^{(d/r)}\big)^{\rm H}$ with $\bm{H}_{0,{\rm d},[~:k^{*}]}^{(d/r)}$ denoting the $k^{*}$th column of $\bm{H}_{0,{\rm d}}^{(d/r)}$, while $\bm{\Psi}_{\widetilde{\bm{H}}_{0,{\rm r},[~:k^{*}]}^{(d/r)}}$ and $\bm{\Psi}_{\widehat{\bm{H}}_{0,{\rm r},[~:k^{*}]}^{(d/r)}}$ follow the definitions of (\ref{eq12}) and (\ref{eq13}) but with $\widetilde{\bm{H}}_{0,{\rm r}}^{(d/r)}$ and $\widehat{\bm{H}}_{0,{\rm r}}^{(d/r)}$ being replaced by $\widetilde{\bm{H}}_{0,{\rm r},[~:k^{*}]}^{(d/r)}$ and $\widehat{\bm{H}}_{0,{\rm r},[~:k^{*}]}^{(d/r)}$ in the associated equations, respectively. Furthermore, the normalization factor $\lambda^{(d/r)}$ is given by
\begin{align}\label{eq27}
\lambda^{(d/r)} = \left(\frac{1}{K}\text{Tr}\left\{\nu_{r}^{2}\bm{B}_{\bm{H}_{0, {\rm d}}^{(d/r)}} + \zeta_{r}^{2}\bm{\Psi}_{\widehat{\bm{H}}_{0, {\rm r}}^{(d/r)}}\right\}\right) ,
\end{align}
in which $\bm{B}_{\bm{H}_{0, {\rm d}}^{(d/r)}} = \bm{vec}(\bm{H}_{0, {\rm d}}^{(d/r)})\bm{vec}(\bm{H}_{0, {\rm d}}^{(d/r)})^{\rm H}$.

The interference plus noise power $\mathcal{E}\left\{P_{{\rm IN},k^{*}}^{(d/r)}\right\}$ can be expressed as Eq.~(\ref{eq28}).

 \begin{figure*}[tp!]\setcounter{equation}{27}
\vspace*{-4mm}
\begin{align}\label{eq28}
\mathcal{E}\left\{P_{{\rm IN},k^{*}}^{(d/r)}\right\} \!\!= & \lambda^{(d/r)}\sigma_{n}^{2}\text{Tr}\left\{\nu_{r}^{2}\bm{B}_{\bm{H}_{0, {\rm d}, [~:k^{*}]}^{(d/r)}}\!\! +\!\! \zeta_{r}^{2}\bm{\Psi}_{\widehat{\bm{H}}_{0, {\rm r}, [~:k^{*}]}^{(d/r)}}\right\} \!\!+ \!\!\lambda^{(d/r)}P_{{\rm rx},0(k^{*})}^{(d,s)}\text{Tr}\left\{\bm{\Psi}_{\widetilde{\bm{H}}_{0, {\rm r}, [~:k^{*}]}^{(d/r)}}\left(\nu_{r}^{2}\bm{B}_{\bm{H}_{0, {\rm d}, [~:k^{*}]}^{(d/r)}} \!\!+ \zeta_{r}^{2}\bm{\Psi}_{\bm{H}_{0, {\rm r}, [~:k^{*}]}^{(d/r)}}\right)\right\} \nonumber\\
&  \!\!+  \!\sum\limits_{k = 1, k \neq k^{*}}^{K} \! \!\lambda^{(d/r)}P_{{\rm rx},0(k)}^{(d,s)}\text{Tr} \!\left\{ \!\left(\nu_{r}^{2}\bm{B}_{\bm{H}_{0, {\rm d}, [~:k^{*}]}^{(d/r)}}  \!+ \! \zeta_{r}^{2}\bm{\Psi}_{\widehat{\bm{H}}_{0, {\rm r}, [~:k^{*}]}^{(d/r)}}\right) \!\left(\nu_{r}^{2}\bm{B}_{\bm{H}_{0, {\rm d}, [~:k]}^{(d/r)}} \! + \! \zeta_{r}^{2}\bm{\Psi}_{\bm{H}_{0, {\rm r}, [~:k]}^{(d/r)}}\right)\right\} \nonumber\\
& \!+ \! \sum\limits_{a=1}^A \!\sum\limits_{k = 1}^{K} \!\lambda^{(d/r)}P_{{\rm rx},a(k)}^{(d,s)}\text{Tr} \!\left\{ \!\left(\nu_{r}^{2}\bm{B}_{\bm{H}_{0, {\rm d}, [~:k^{*}]}^{(d/r)}} \! +  \!\zeta_{r}^{2}\bm{\Psi}_{\widehat{\bm{H}}_{0, {\rm r}, [~:k^{*}]}^{(d/r)}}\right) \!\left(\nu_{r}^{2}\bm{B}_{\bm{H}_{a, {\rm d}, [~:k]}^{(d/r)}} \! +  \!\zeta_{r}^{2}\bm{\Psi}_{\bm{H}_{a, {\rm r}, [~:k]}^{(d/r)}}\right)\right\} .
\end{align}
\setcounter{equation}{30}
\begin{align}\label{eq31}
 \bar{P}_{{\rm IN},k^*}^{(r/g)} =& \lambda^{(r/g)}\sigma_{v}^{2}\text{Tr}\left\{\nu_{g}^{2}\bm{B}_{\bm{H}_{0, {\rm d}, [~:k^{*}]}^{(r/g)}} + \zeta_{g}^{2}\bm{\Psi}_{\widehat{\bm{H}}_{0, {\rm r}, [~:k^{*}]}^{(r/g)}}\right\} + \lambda^{(r/g)}P_{{\rm rx},0(k^{*})}^{(r,s)}\text{Tr}\left\{\bm{\Psi}_{\widetilde{\bm{H}}_{0, {\rm r}, [~:k^{*}]}^{(r/g)}}\left(\nu_{g}^{2}\bm{B}_{\bm{H}_{0, {\rm d}, [~:k^{*}]}^{(r/g)}} + \zeta_{g}^{2}\bm{\Psi}_{\bm{H}_{0, {\rm r}, [~:k^{*}]}^{(r/g)}}\right)\right\} \nonumber\\
&  \!\!+  \!\sum\limits_{k = 1, k \neq k^{*}}^{K} \! \!\lambda^{(r/g)}P_{{\rm rx},0(k)}^{(r,s)}\text{Tr} \!\left\{ \!\left(\nu_{g}^{2}\bm{B}_{\bm{H}_{0, {\rm d}, [~:k^{*}]}^{(r/g)}}  \!+ \! \zeta_{g}^{2}\bm{\Psi}_{\widehat{\bm{H}}_{0, {\rm r}, [~:k^{*}]}^{(r/g)}}\right) \!\left(\nu_{g}^{2}\bm{B}_{\bm{H}_{0, {\rm d}, [~:k]}^{(r/g)}} \! + \! \zeta_{g}^{2}\bm{\Psi}_{\bm{H}_{0, {\rm r}, [~:k]}^{(r/g)}}\right)\right\} \nonumber \\
& \!+ \! \sum\limits_{b=1}^{B-1} \!\sum\limits_{k = 1}^{K} \!\lambda^{(r/g)}P_{{\rm rx},b(k)}^{(r,s)}\text{Tr} \!\left\{ \!\left(\nu_{g}^{2}\bm{B}_{\bm{H}_{0, {\rm d}, [~:k^{*}]}^{(r/g)}} \! +  \!\zeta_{g}^{2}\bm{\Psi}_{\widehat{\bm{H}}_{0, {\rm r}, [~:k^{*}]}^{(r/g)}}\right) \!\left(\nu_{g}^{2}\bm{B}_{\bm{H}_{b, {\rm d}, [~:k]}^{(r/g)}} \! +  \!\zeta_{g}^{2}\bm{\Psi}_{\bm{H}_{b,{\rm r}, [~:k]}^{(r/g)}}\right)\right\} . 
\end{align}
\hrulefill
\vspace*{-4mm}
\end{figure*}

\subsection{The achievable throughput of the RUAV-to-GS link}\label{S3.2}

Similarly, the achievable throughput of the $k^{*}$th DTA in the targeted RUAV-to-GS link is given by 
\setcounter{equation}{28}
\begin{align}\label{eq29}
C_{k^*}^{(r/g)} \approx \log_{2}\left(1 + \frac{\bar{P}_{\text{S},k^*}^{(r/g)}}{\bar{P}_{\text{IN},k^*}^{(r/g)}}\right) ,
\end{align}
where the signal power $\bar{P}_{{\rm S},k^*}^{(r/g)}=\mathcal{E}\left\{P_{{\rm S},k^*}^{(r/g)}\right\}$ and
 the interference plus noise power $\bar{P}_{{\rm IN},k^*}^{(r/g)}=\mathcal{E}\left\{P_{{\rm IN},k^*}^{(r/g)}\right\}$ are given respectively by Eq.~(\ref{eq30}) and Eq.~(\ref{eq31}), respectively. 
\begin{align}\label{eq30}
\bar{P}_{{\rm S},k^*}^{(r/g)} =& \lambda^{(r/g)}P_{{\rm rx},0}^{(r,s)}\bigg(\text{Tr}\bigg\{\nu_{g}^{2}\bm{B}_{\bm{H}_{0, {\rm d}, [~:k^{*}]}^{(r/g)}} + \zeta_{g}^{2}\bm{\Psi}_{\widehat{\bm{H}}_{0, {\rm r}, [~:k^{*}]}^{(r/g)}} \nonumber \\
& + \zeta_{g}^{2}\bm{\Psi}_{\widehat{\bm{H}}_{0, {\rm r}, [~:k^{*}]}^{(r/g)}}\bm{\Psi}_{\widetilde{\bm{H}}_{0, {\rm r}, [~:k^{*}]}^{(r/g)}}\bigg\}\bigg)^{2} ,
\end{align}

\subsection{Distance-based ACM}\label{S3.3}

ACM is a powerful link adaptation technique conceived for improving bandwidth efficiency (BE), which was traditionally developed in line with the instantaneous SINR. However, upon invoking (\ref{eq2}) and (\ref{eq3}), it can be readily seen that the received signal power decreases upon increasing the communication distance, which in turn reduces both the uplink SINR and the achievable uplink throughput. By considering the large-scale geographic distribution of aeronautical communications, we follow the philosophy of distance-based ACM proposed in our earlier work \cite{zhang2018adaptive}. In the following, we take the DCDS-to-RUAV link as an  example for briefly introducing the distance-based ACM, whilst the operations of the distance-based ACM assisted RUAV-to-GS link adaptation follows the same procedure. Note that the RUAV-to-GS link has the same system parameters, including the parameters of distance-based ACM, as the DCDS-to-RUAV link.

Given the total bandwidth $B_{\text{total}}$, the number of subcarriers $N_c$, the number of CP $N_{\text{cp}}$ and the number of ACM modes $Q$ associated with a set of distance-based switching thresholds $\{d_{m}^{(d/r)}\}_{q = 0}^{Q}$,  we have $d_{0}^{(d/r)}\! =\! D_{\text{max}}^{(d/r)}$ and $d_{Q}^{(d/r)}\! =\! D_{\text{min}}^{(d/r)}$, where $D_{\text{max}}^{(d/r)}$ and $D_{\text{min}}^{(d/r)}$ are the minimum safe separation distance and the maximum communication distance. The key operations of the MF-based RC scheme and distance-based ACM are summarized as follows.
\begin{itemize}
\item [1)] {\bf{Position broadcasting}}. The RUAV broadcasts its position to the DCDS.

\item [2)] {\bf{Pilot training}}. The RUAV estimates the channel matrix $\bm{H}_{0}^{(d/r)}$ based on the pilot symbols sent by the DCDS.

\item [3)] {\bf{ACM mode selection}}. The $k$th drone of the desired drone swarm chooses its ACM mode based on its distance from the RUAV, $d_{0,k}^{(d/r)}$, according to
\setcounter{equation}{31}
\begin{align}\label{eq32}
  \text{If } d_{q}^{(d/r)} \le d_{0,k}^{(d/r)} < d_{q-1}^{(d/r)}:& \text{ choose mode } q , \, \nonumber \\& q \in \{1,2,\cdots , Q\} .
\end{align}

\item [4)] {\bf{Data transmission}}. Each drone transmits its signal using the ACM mode chosen according to (\ref{eq32}). The data rate of the $k$th drone is given by
\begin{align}\label{eq33}
R_{\text{total}, k}^{(d/r)} =& B_{\text{total}}r_{c,q}\log_{2}\left(M_{q}\right)\frac{N_{c}}{N_{c} + N_{\text{cp}}} ,
\end{align}
where $r_{c,q}$ and $M_{q}$ are the $q$th ACM mode's coding rate and modulation order, respectively. 

\item [5)] {\bf{Data reception}}. The RUAV detects the DCDS's signal by applying the MF-based  RC matrix $\bm{W}_{0}^{(d/r)}$ to the received signal.
\end{itemize}

\section{Optimization problems of relaying-assisted FANET}\label{S4}

We consider a challenging scenario in that the end-to-end communication requires the assistance of an RUAV, because there is no direct communication link between the DCDS and the GS. The DCDS is capable of communicating with the remote GS either continuously or intermittently with the assistance of a RUAV as illustrated in Fig.~\ref{FIG1}, depending on the end-to-end distance from the DCDS to the GS. Explicitly, let us denote the maximum communication distances of the DCDS-to-RUAV link and the RUAV-to-GS link as $D_{\text{max}}^{(d/r)}$ and $D_{\text{max}}^{(r/g)}$, respectively, whilst the minimum communication distances of the DCDS-to-RUAV link and the RUAV-to-GS link are denoted as $D_{\text{min}}^{(d/r)}$ and $D_{\text{min}}^{(r/g)}$, respectively. The total end-to-end distance from the DCDS to the GS is denoted by $D^{(d/g)}$. Let us assume that the drones in a swarm have the same distance from the RUAV\footnote{This assumption approximately holds, since the drones in a swarm are relatively close to each other and the distances between the drones in a DCDS are much smaller compared to the distance between the DCDS and RUAV. Moreover, this assumption may also be made to hold by coordinately deploying the drones in a swarm, which are controllable by the RUAV for cooperation.}. Thus, we have $d_{0, 1}^{(d/r)} = d_{0, 2}^{(d/r)} = \cdots = d_{0, K}^{(d/r)} = d_{0}^{(d/r)}$, where again $d_{0, k}^{(d/r)}$ is the communication distance between the $k$th drone and the RUAV.  The RUAV starts from the point having the minimum distance from the DCDS, say $A_{0}$, where we have $d_{0}^{(d/r)} = d_{0}^{(d/r)} (0) = D_{\text{min}}^{(d/r)}, k = 1, 2, \cdots, K$. 

There are two scenarios to be investigated depending on the relationship between $\big(D_{\text{max}}^{(d/r)} + D_{\text{max}}^{(r/g)}\big)$ and $D^{(d/g)}$, when a relay node is needed, namely that of a {\bf{static relay}} when $D_{\text{max}}^{(d/r)} < D^{(d/g)} \le \big(D_{\text{max}}^{(d/r)} + D_{\text{max}}^{(r/g)}\big)$ and that of a {\bf{mobile relay}} when $\big(D_{\text{max}}^{(d/r)} + D_{\text{max}}^{(r/g)}\big) < D^{(d/g)}$. 

\subsection{Static relay}\label{S4-1}

When $D_{\text{max}}^{(d/r)} < D^{(d/g)} \le \big(D_{\text{max}}^{(d/r)} + D_{\text{max}}^{(r/g)}\big)$, there exist simultaneous communication links for both the DCDS-to-RUAV and the RUAV-to-GS. In this case, the RUAV may act as a static relay between the GS and the DCDS. OFDMA is used by the RUAV for receiving and forwarding the DCDS's data, and the delay imposed in the static-relay scenario is $\tau_{s} = 0$, corresponding to omitting the delay of signal processing associated both with reception and forwarding.  However, the position of the RUAV can be optimized in order to maximize the achievable end-to-end throughput. Explicitly, this optimization problem is formulated as
\begin{align}\label{eq34}
\text{Find: } & \bm{d}_{\text{opt}} = (D^{(d/r)}, D^{(r/g)}) \\
\label{eq35}
\text{to maximize: } & R_{e,\text{sum}} = \sum\limits_{k = 1}^{K} R_{e,k} , \\
\label{eq36}
\text{subject to: } & \left\{\begin{array}{l} 
  D^{(d/r)} + D^{(r/g)} = D^{(d/g)} ,\\
  D_{\text{min}}^{(d/r)} \le D^{(d/r)} \le D_{\text{max}}^{(d/r)} , \\
  D_{\text{min}}^{(r/g)} \le D^{(r/g)} \le D_{\text{max}}^{(r/g)} ,
\end{array}	
\right.									 
\end{align}
where $R_{e,k} = \min \, \left\{R_{\text{total}, k}^{(d/r)}\left(D^{(d/r)}\right), R_{\text{total}, k}^{(r/g)}\left(D^{(r/g)}\right)\right\}$ is the $k$th drone's effective end-to-end rate, $R_{\text{total}, k}^{(d/r)}\left(D^{(d/r)}\right)$ and $R_{\text{total}, k}^{(r/g)}\left(D^{(r/g)}\right)$ are the DCDS-to-RUAV link data rate corresponding to $D^{(d/r)}$ and the RUAV-to-GS link data rate corresponding to $D^{(r/g)}$, respectively.

\begin{figure}[tp!]
\vspace*{-3mm}
\begin{center}
 \includegraphics[width=0.5\textwidth]{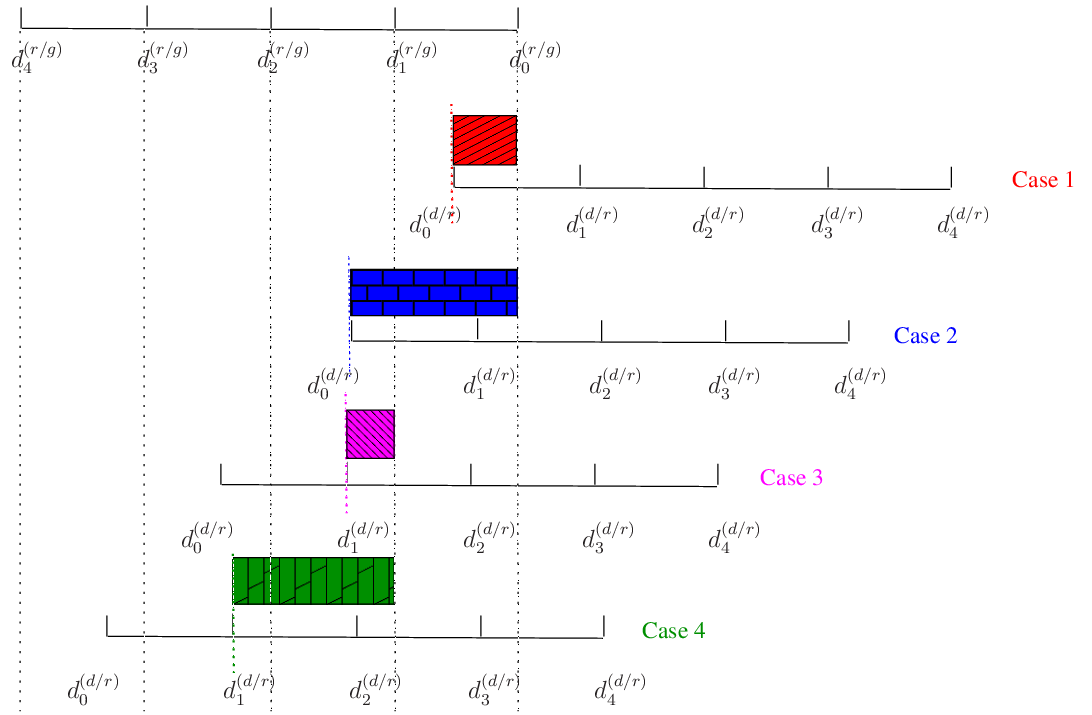}
\end{center}
\vspace*{-2mm}
\caption{An illustration of RUAV's optimal position.}
\label{FIG2}
\vspace*{-5mm}
\end{figure}

In order to identify both the achievable maximum end-to-end throughput and the optimal position of the RUAV, we define $D_{o} = D_{\text{max}}^{(d/r)} + D_{\text{max}}^{(r/g)} - D^{(d/g)}$ as the overlapped communication range of the DCDS and the GS, as seen in Fig.~\ref{FIG2}. The maximum achievable end-to-end throughput and the corresponding position of the RUAV can be found as follows.

Given $D_{o}$, we can readily identify that the RUAV's ACM mode switching thresholds $d_{0}^{(r/g)}$, $d_{1}^{(r/g)},\cdots, d_{I}^{(r/g)}$ are located within the overlapped communication range of the DCDS and the GS, whilst the DCDS's ACM mode switching thresholds $d_{0}^{(d/r)}, d_{1}^{(d/r)},\cdots, d_{J}^{(d/r)}$ are located within the overlapped communication range of the DCDS and the GS. Because the ACM mode is changed upon crossing a specific switching threshold, we can evaluate the achievable throughput at the points of $(d_{(l)}^{(r/g)},d_{(l)}^{(d/r)})$, $l\! =\! 1, 2, \cdots, L$ and $L\! =\! I + J - O$. Here, $O$ is the number of  points for any $i\! =\! 1, 2, \cdots, I$ and $j\! =\! 1, 2, \cdots, J$, $d_{i}^{(r/g)}\! =\! D^{(d/g)} - d_{j}^{(d/r)}$. Furthermore, $d_{(l)}^{(r/g)}$ is a value selected from $\Big\{d_{0}^{(r/g)},  d_{1}^{(r/g)},\cdots, d_{I}^{(r/g)}, D^{(d/g)} - d_{0}^{(d/r)}, D^{(d/g)} - d_{1}^{(d/r)}, \cdots, D^{(d/g)} - d_{J}^{(d/r)}\Big\}$ and we sort $d_{(l)}^{(r/g)}$, $l\! =\! 1, 2, \cdots, L$, in the ascending order of $d_{(1)}^{(r/g)} < d_{(2)}^{(r/g)} <\cdots <d_{(L)}^{(r/g)}$. The maximum achievable end-to-end throughput can be expressed as follows
\begin{align}\label{eq37}
R_{e,\text{sum}, \text{max}}\!\! =\!\! \max \! \left\{\!\sum\limits_{k = 1}^{K} R_{e,k}\left(d_{(l)}^{(r/g)},d_{(l)}^{(d/r)}\right)\!\right\}, l = 1, 2, \cdots, L .
\end{align}
The case of a single point to achieve the maximum throughput only happens when $D^{(d/g)}\! =\! d_{0}^{(r/g)} + d_{0}^{(d/r)}$. Otherwise there will be at least two critical points, within which it will be possible to achieve the maximum throughput, as indicated by the shadow areas in Fig.~\ref{FIG2}. Let us denote these critical points by $\big(d_{(l^{*})}^{(r/g)},d_{(l^{*})}^{(d/r)}\big)$, $l\! =\! 1, 2, \cdots, L^{*}$, where $d_{(l^{*})}^{(r/g)}$ and $d_{(l^{*})}^{(d/r)}$, $l^{*}\! =\! 1, 2, \cdots, L^{*}$, have been sorted as $d_{(1)}^{(r/g)} < d_{(2)}^{(r/g)} < \cdots < d_{(L^{*})}^{(r/g)}$ and $d_{(1)}^{(d/r)} < d_{(2)}^{(d/r)} < \cdots < d_{(L^{*})}^{(d/r)}$, respectively. Then, the RUAV is capable of achieving the maximum throughput when its distance from the GS is in the range of $\Big[d_{(l^{*})}^{(r/g)},d_{(l^{*})}^{(d/r)}\Big]$, as shown by the different patterns in Fig.~\ref{FIG2}.  

\begin{figure}[tbp!]
\vspace*{-4mm}
\begin{center}
 \includegraphics[width=0.5\textwidth]{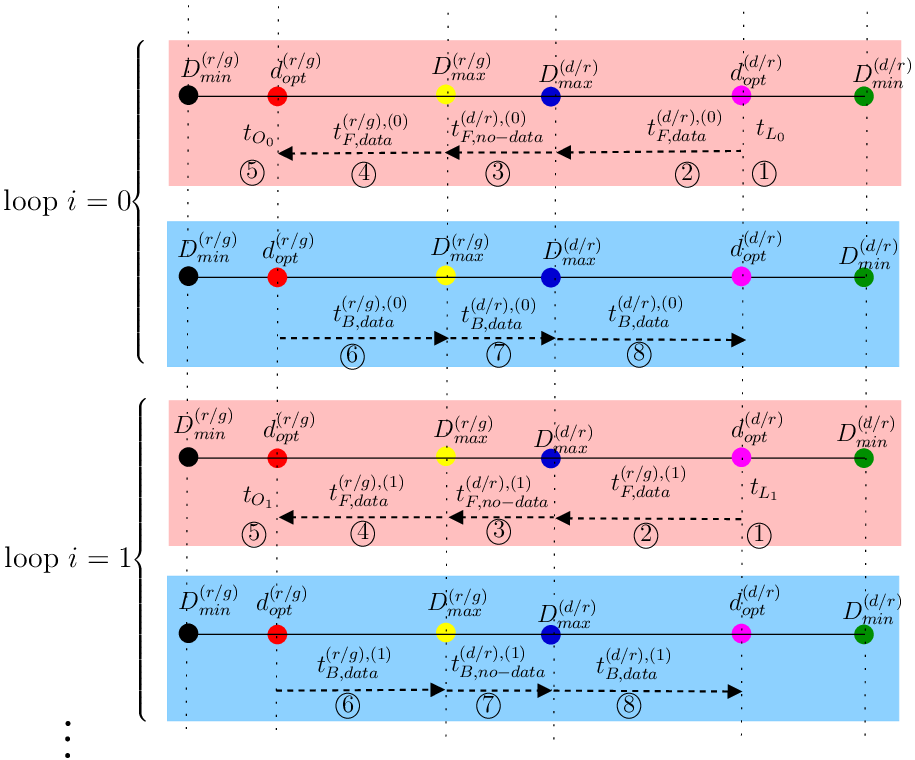}
\end{center}
\vspace*{-2mm}
\caption{State transition of the mobile relaying.}
\label{FIG3}
\vspace*{-2mm}
\end{figure}

\subsection{Mobile relay for data ferry}\label{S4-2}

Given the maximum RUAV to GS communication distance
$D_{\text{max}}^{(r/g)}$ and RUAV to DCDS distance
$D_{\text{max}}^{(d/r)}$, if we have $\big(D_{\text{max}}^{(d/r)} +
D_{\text{max}}^{(r/g)}\big) < D^{(d/g)}$, the RUAV can only access
either the GS or DCDS, or in fact neither of them. So, the
DCDS-to-RUAV and RUAV-to-GS links do not exist concurrently. In this
scenario, the traditional static relaying scheme of
\emph{Section~\ref{S4-2}} will no longer work. In order to relay data
from the DCDS to the GS, the RUAV first has to acquire the data at a
DCDS point and then fly to a GS to offload it, as illustrated in
Fig.~\ref{FIG1}. Intuitively, a buffer is required by the RUAV for
storing data before ferrying it to the GS. We assume that the buffer
size is $T_{b}$ Gigabit (GB). Once the RUVA offloaded the data to the
GS, it will return to the DCDS point to acquire more data, which
completes a whole loop, as shown in Fig.~\ref{FIG3}. Explicitly, there
are eight states for a complete loop.  Let us assume that the
procedure will be commenced from a DCDS point for data off-loading,
and the ensuing state transitions will be detailed in the rest of this
subsection by frequently referring to Fig.~\ref{FIG3}.

\begin{itemize}
\item [$\circled{1}$] {\bf{Data loading in the Vicinity of a DCDS}}.
In this state, the RUVA hovers at a DCDS for acquiring data from it.
The instantaneous data stored in the RUAV's buffer $T_{d}$ at time $t$ in second (s) is given by
\begin{align}\label{eq38}
T_{d}(t) \!\!= \!\!T_{d}(t - 1)\!\! +\!\! R_{\text{total}, k}^{(d/r)}\left(D_{i}^{(d/r)}\right) , ~ t_{S_{1, i}} \!\!< \!\!t \le t_{S_{1, i}} \!\!+ \!\!t_{L_{i}} ,
\end{align}
where $D_{i}^{(d/r)}\! =\! D_{\text{opt}}^{(d/r)}$, and $T_{d}(0)\!
=\! 0$, since no data is stored in the buffer at time $t\! =\! 0$. The
RUAV will then fly to the GS, when the buffer-fullness reaches the
upper threshold $T_{th}^{up}$, so the data loading duration $t_{L_{i}}$ at
the $i$th loop is given by
\begin{align}\label{eq39}
t_{L_{i}} = \left\lfloor \frac{T_{th}^{up} - T_{d}(t_{S_{1, i}})}{R_{\text{total}, k}^{(d/r)}\left(D_{i}^{(d/r)}\right)} \right\rfloor ,
\end{align}
where again, $T_{th}^{up}\! =\! \alpha T_b$ is the upper threshold for
loading data at a DCDS with $0 < \alpha \le 1$ being the
maximum factor of caching data, and $t_{S_{1, i}}$ is given by
\begin{align}\label{eq40}
t_{S_{1, i}}  =  & \left\{
\begin{array}{cl}
0 , &   i = 0 , \\
t_{S_{1, i - 1}}	+  t_{P_{i - 1}} , &   i \ge 1 ,
\end{array}
\right.
\end{align}
in which $t_{P_{i}} = t_{L_{i}} + t_{F,data}^{(d/r),(i)} +  t_{F,no-data}^{(d/r),(i)} + t_{F,data}^{(r/g),(i)} +  t_{O_{i}} + t_{B,data}^{(r/g),(i)} + t_{B,no-data}^{(d/r),(i)} + t_{B,data}^{(d/r),(i)}$ is the period of the RUAV cycling from states $\circled{1}$  to $\circled{8}$ in Fig.~\ref{FIG3}. Note that we have $t_{P_{i}} = t_{L_{0}}$. The definitions of the time periods, $t_{F,data}^{(d/r),(i)}$, $t_{F,no-data}^{(d/r),(i)}$, $t_{F,data}^{(r/g),(i)}$, $t_{B,data}^{(r/g),(i)}$, $t_{B,no-data}^{(d/r),(i)}$ and $t_{B,data}^{(d/r),(i)}$, can be found in Fig.~\ref{FIG3}.

Moreover, the communication distance between the DCDS and the RUAV remains $D_{i}^{(d/r)}$, i.e., $d_{0}^{(d/r)} (t) = D_{i}^{(d/r)}$ for $t_{S_{1, i}} + 1 \le t \le t_{S_{1, i}} + t_{L_{i}}$.  The data cumulatively received by the GS is given by
\begin{align}\label{eq41}
T_{r}(t) =& T_{r}(t - 1), ~ t_{S_{1, i}} < t \le t_{S_{1, i}} + t_{L_{i}} ,
\end{align}
with $T_{r}(0) = 0$. 

\item [$\circled{2}$] {\bf{Flying towards the GS whilst continuing data loading, when the RUAV is  within the maximum communication range of the DCDS}}.

During this specific state of Fig.~\ref{FIG3}, the RUAV is within the
communication range of the DCDS. Hence, the RUAV continues to load
the data whilst flying towards the GS. The distance between the RUAV
and DCDS at instant $t$ is given by
\begin{align}\label{eq42}
d_{0}^{(d/r)} (t) \!\!=\!\! d_{0}^{(d/r)} (t - 1) \!\!+\!\! V , ~ t_{S_{2, i}} \!\!<\!\! t \le t_{S_{2, i}} + t_{F,data}^{(d/r), (i)} ,
\end{align}
where  $V$ is the RUAV's flying velocity in m/s and $t_{S_{2, i}}$ is given by
\begin{align}\label{eq43}
t_{S_{2, i}} = t_{S_{1, i}} + t_{L_{i}} , ~ i = 0, 1, 2, \cdots .
\end{align}
The data stored in the RUAV's buffer $T_{d}$ at instant $t$ is given by
\begin{align}\label{eq44}
T_{d}(t) =& T_{d}(t - 1) + R_{\text{total}, k}^{(d/r)}\left(d_{0}^{(d/r)} (t)\right) , \nonumber\\ & t_{S_{2, i}} < t \le t_{S_{2, i}} + t_{F,data}^{(d/r), (i)} .
\end{align}

The data received cumulatively by the GS is given by
\begin{align}\label{eq45}
T_{r}(t) =& T_{r}(t - 1), ~ t_{S_{2, i}} < t \le t_{S_{2, i}} + t_{F,data}^{(d/r), (i)} .
\end{align}

\item [$\circled{3}$] {\bf{Flying towards the GS but the RUAV is out of both the DCDS's and the GS's communication range}}.

In this state of Fig.~\ref{FIG3}, there is no DCDS-to-RUAV and
RUAV-to-GS communication links, since the RUAV is out of both the DCDS's
and the GS's communication range. Hence, the RUAV's buffer remains
unchanged during this state. Explicitly, the data stored in the RUAV's
buffer and the distance between the RUAV and the DCDS are given
respectively by
\begin{align}\label{eq46}
T_{d}(t) =& T_{d}(t - 1), ~ t_{S_{3, i}} < t \le t_{S_{3, i}} + t_{F,no\text{-}data}^{(i)} , \\
\label{eq47}
d_{0}^{(d/r)} (t)\!\! =& d_{0}^{(d/r)} (t - 1) \!\!+ V , ~ t_{S_{3, i}}\!\! < \!\!t \!\!\le\!\! t_{S_{3, i}} + t_{F,no\text{-}data}^{(i)} ,
\end{align}
where $t_{S_{3, i}}$ is given as
\begin{align}\label{eq48}
t_{S_{3, i}} =    t_{S_{2, i}} + t_{F,data}^{(d/r), (i)} , ~ i = 0, 1, 2, \cdots .
\end{align}

The data received cumulatively by the GS remains unchanged as well, which is given by
\begin{align}\label{eq49}
T_{r}(t) =& T_{r}(t - 1), ~ t_{S_{3, i}} < t \le t_{S_{3, i}} + t_{F,no\text{-}data}^{(i)} .
\end{align}

\item [$\circled{4}$] {\bf{Flying towards the GS whilst offloading data to the GS, when the RUAV is within the communication range of the GS}}.

In this state, the RUAV is within the communication range of the GS, but it
is out of the communication range of the DCDS, as seen in
Fig.~\ref{FIG3}. So, the RUAV begins to offload its data to the
GS. The instantaneous distance between the RUAV as well as the GS and
the data stored in the RUAV's buffer are given by
 \begin{align}\label{eq50}
d_{0}^{(r/g)} (t) \!\!=& d_{0}^{(r/g)} (t - 1)\!\! - \!\!V , ~ t_{S_{4, i}} \!\!<\!\! t \!\!\le\!\! t_{S_{4, i}} + t_{F,data}^{(r/g), (i)} , \\
\label{eq51}
T_{d}(t) =& T_{d}(t - 1) - R_{\text{total}, k}^{(r/g)}\left(d_{0}^{(r/g)} (t)\right), \nonumber\\ &~ t_{S_{4, i}} < t \le t_{S_{4, i}} + t_{F,data}^{(r/g), (i)} , 
\end{align}
respectively, where $d_{0}^{(r/g)} (t_{S_{4, i}}) = D_{\text{max}}^{(r/g)}$ and $t_{S_{4, i}}$ is given by
\begin{align}\label{eq52}
t_{S_{4, i}} =    t_{S_{3, i}} + t_{F,no\text{-}data}^{(d/r), (i)} , ~ i = 0, 1, 2, \cdots .
\end{align}

The accumulated data received by the GS is given by
\begin{align}\label{eq53}
T_{r}(t) =& T_{r}(t - 1) + R_{\text{total}, k}^{(r/g)}\left(d_{0}^{(r/g)} (t)\right), \nonumber\\
&~ t_{S_{4, i}} < t \le t_{S_{4, i}} + t_{F,data}^{(r/g), (i)} .
\end{align}

\item [$\circled{5}$] {\bf{Data offloading in the vicinity of the
    GS}}.  When the RUAV arrives at the optimized near-GS point as
  seen in Fig.~\ref{FIG3}, it will hover at the optimized near-GS
  point, while offloading data to the GS. However, the distance between the
  RUAV and the GS remains unchanged. Hence, the distance between the
  RUAV as well as the GS and the data stored in the RUAV's buffer are given by
 \begin{align}\label{eq54}
&d_{0}^{(r/g)} (t) = D_{\text{opt}}^{(r/g)} , ~ t_{S_{5, i}} < t \le t_{S_{5, i}} + t_{O_{i}} , \\
\label{eq55}
&T_{d}(t) = T_{d}(t - 1) - R_{\text{total}, k}^{(r/g)}\left(D_{\text{opt}}^{(r/g)}\right), \nonumber\\ & \quad\quad\quad\quad t_{S_{5, i}} < t \le t_{S_{5, i}} + t_{O_{i}} , 
\end{align}
respectively, where $t_{S_{5, i}}$ is given by
\begin{align}\label{eq56}
t_{S_{5, i}} =    t_{S_{4, i}} + t_{F,data}^{(r/g), (i)} , ~ i = 0, 1, 2, \cdots .
\end{align}
Furthermore, $t_{O_{i}}$ is formulated as 
\begin{align}\label{eq57}
t_{O_{i}} = \left\lfloor \frac{T_{d}(t_{S_{5, i}}) - T_{th}^{low}}{R_{\text{total}, k}^{(d/r)}\left(D_{\text{opt}}^{(r/g)}\right)} \right\rfloor ,
\end{align}
where $T_{th}^{low}\! =\! \beta T_b$ is the lower threshold for
offloading data at the GS, with $\beta$ being the minimum
factor of offloading data\footnote{As a specific example, given $T_b =
  32$GB and $\beta = 0.1$, when the remaining data in the RUAV's
  buffer is 3.2GB, the RUAV will proceed to the next phase
  $\circled{6}$. }.

The data cumulatively received by the GS is given by
\begin{align}\label{eq58}
T_{r}(t) \!\!=\!\! T_{r}(t - 1) \!\!+\!\! R_{\text{total}, k}^{(r/g)}\left(D_{\text{opt}}^{(r/g)}\right), t_{S_{5, i}} \!\!<\!\! t \!\!\le\!\! t_{S_{5, i}} + t_{O_{i}} .
\end{align}

\item [$\circled{6}$] {\bf{Flying towards the DCDS, whilst offloading data to the GS, when the RUAV is still within the communication range of the GS}}.

At this state of Fig.~\ref{FIG3}, the RUAV flies towards the DCDS, whilst
it continues offloading the remaining data to the GS, since it remains
within the communication range of GS. The distance between the RUAV
and the GS can be expressed as
\begin{align}\label{eq59}
d_{0}^{(r/g)} (t) \!\!=\!\! d_{0}^{(r/g)} (t - 1) \!\!+\!\! V , t_{S_{6, i}} < t \le t_{S_{6, i}} + t_{B,data}^{(r/g), (i)} ,
\end{align}
where  $d_{0}^{(r/g)} (t_{S_{6, i}}) = D_{\text{opt}}^{(r/g)}$ and $t_{S_{6, i}}$ is given by
\begin{align}\label{eq60}
t_{S_{6, i}} = t_{S_{5, i}} + t_{O_{i}} , ~ i = 0, 1, 2, \cdots .
\end{align}
The amount of data $T_{d}(t)$ at instant $t$ stored in the RUAV's buffer is given by
\begin{align}\label{eq61}
T_{d}(t) =& T_{d}(t - 1) - R_{\text{total}, k}^{(r/g)}\left(d_{0}^{(r/g)} (t)\right) , \nonumber \\ & t_{S_{6, i}} < t \le t_{S_{6, i}} + t_{B,data}^{(r/g), (i)} ,
\end{align}
while the data cumulatively received by the GS is given by
\begin{align}\label{eq62}
T_{r}(t) =& T_{r}(t - 1) + R_{\text{total}, k}^{(r/g)}\left(d_{0}^{(r/g)} (t)\right) ,  \nonumber \\ &  t_{S_{6, i}} < t \le t_{S_{6, i}} + t_{B,data}^{(r/g), (i)} .
\end{align}

\item [$\circled{7}$] {\bf{The RUAV flies towards the DCDS but is out of both DCDS's and GS's range}}.

Similar to state $\circled{3}$ of Fig.~\ref{FIG3}, because the RUAV is
out of both the DCDS's and GS's communication range in this state, 
there is no data transmission and data reception. The RUAV's buffer
remains unchanged. Explicitly, the data stored in the RUAV's buffer
and the distance between the RUAV and the DCDS is given by
 \begin{align}\label{eq63}
T_{d}(t) =& T_{d}(t - 1), ~ t_{S_{7, i}} < t \le t_{S_{7, i}} + t_{B,no\text{-}data}^{(i)} , \\
\label{eq64}
d_{0}^{(r/g)} (t) \!\!=& \!\!d_{0}^{(r/g)} (t - 1)\!\! +\!\! V , ~ t_{S_{7, i}} \!\!<\!\! t \le t_{S_{7, i}} + t_{B,no\text{-}data}^{(i)} ,
\end{align}
respectively, where $t_{S_{7, i}}$ is formulated as
\begin{align}\label{eq65}
t_{S_{7, i}} = t_{S_{6, i}} + t_{B,data}^{(r/g), (i)} , ~ i = 0, 1, 2, \cdots .
\end{align}

The data cumulatively received by the GS remains unchanged as well,
which is given by
\begin{align}\label{eq66}
T_{r}(t) =& T_{r}(t - 1) , ~ t_{S_{7, i}} < t \le t_{S_{7, i}} + t_{B,no\text{-}data}^{(i)} .
\end{align}

\item [$\circled{8}$] {\bf{The RUAV flies towards the DCDS and starts to load data when the RUAV is within the communication range of the DCDS}}.

When the RUAV passes the point of $D_{\text{max}}^{(d/r)}$, it will be
within the communication range of the DCDS. Then the RUAV will
automatically load data from the DCDS into its buffer. The instantaneous
distance between the RUAV and the DCDS can be formulated as
\begin{align}\label{eq67}
d_{0}^{(d/r)} (t) \!\!=\!\! d_{0}^{(d/r)} (t - 1) \!\!-\!\! V , ~ t_{S_{8, i}} \!\!< \!\!t \le t_{S_{8, i}} + t_{B,data}^{(d/r), (i)} ,
\end{align}
where  $d_{0}^{(d/r)} (t_{S_{8, i}}) = D_{\text{max}}^{(d/r)}$ and $t_{S_{8, i}}$ is given by
\begin{align}\label{eq68}
t_{S_{8, i}} = t_{S_{7, i}} + t_{B,no\text{-}data}^{(i)} , ~ i = 0, 1, 2, \cdots .
\end{align}

The data $T_{d}$ stored in the RUAV's buffer at time $t$ is given by
\begin{align}\label{eq69}
T_{d}(t) =& T_{d}(t - 1) + R_{\text{total}, k}^{(d/r)}\left(d_{0}^{(d/r)} (t)\right) , \nonumber \\ & t_{S_{8, i}} < t \le t_{S_{8, i}} + t_{B,data}^{(d/r), (i)} .
\end{align}

The data accumulated by the GS remains unchanged, as formulated in Eq.~(eq70), since the RUAV is out of the communication range of the GS. 
\begin{align}\label{eq70}
T_{r}(t) =& T_{r}(t - 1) , ~ t_{S_{8, i}} < t \le t_{S_{8, i}} + t_{B,data}^{(d/r), (i)} .
\end{align}
\end{itemize}

As discussed above, the RUAV periodically loads the data from the DCDS
and offloads the data to the GS when it is flying back and forth between
the DCDS and the GS. Let us define the \emph{effective end-to-end
  connection delay} as the time between the instant when the GS begins
to receive the DCDS's data relayed by the RUAV and the instant when
the RUAV flies over the point $B_{\text{max}}$ at $\text{loop} = 0$,
which is given by
\begin{align}\label{eq71}
\tau_{0} = & t_{L_{0}} + t_{F,data}^{(d/r), (0)} + t_{F,no\text{-}data}^{(0)} .
\end{align}
Let us define the \emph{effective end-to-end average data rate}
$R_{e}(t)$ at time $t$ as the ratio of the accumulated transmitted
data volume over the period of time considered, i.e.,
\begin{align}\label{eq72}
R_{e}(t) = \frac{T_{r}(t)}{t} ,
\end{align}
which also represents the end-to-end data rate experienced at time $t$ by the GS.

There is no data received by the GS at states $\circled{1}$,
$\circled{2}$, $\circled{3}$, $\circled{7}$ and $\circled{8}$ of
Fig.~\ref{FIG3}, and the minimum values of the \emph{effective
  end-to-end average data rate} curve appear at $t_{S_{4, i}}$, $i\!
=\! 1, 2, \cdots$ when the state changes from $\circled{3}$ to
$\circled{4}$. Furthermore, $R_{e}(t_{S_{4, 1}}) < R_{e}(t_{S_{4, 2}})
< \cdots$ for $i\! =\! 1, 2, \cdots$. Let us define $t^*$ as the delay
imposed, when meeting a minimum effective end-to-end average
data rate $R_e^*$. Depending on the particular value of $R_e^*$ in
comparison to $R_{e}(t_{S_{4,i}})$, $t^*$ can be determined as
\begin{align}\label{eq73}
 t^* = t : ~ R_e(t)= & R_e^* \text{ with } t_{S_{4, i - 1}} < t <  t_{S_{7, i - 1}} \nonumber \\ &\text{ if } R_{e}(t_{S_{4, i}}) \le R_e^* < R_{e}(t_{S_{4, i+1}}) .
\end{align}
The optimization problem is to find the optimal near-DCDS loading
point at $d_{\rm opt}^{(d/r)}$ and the optimal near-GS offloading
point $d_{\rm opt}^{(r/g)}$ as well as the optimal factor of caching
data $\alpha_{\rm opt}$ and the optimal factor of offloading data
$\beta_{\rm opt}$ for maximizing a given effective minimum end-to-end
average data rate $R_e^*$, whilst minimizing the delay $t^*$
imposed. Without loss of generality, let us maximize $R_e^*$, while
simultaneously minimizing the delay $t^{*}$ imposed. Explicitly, the
resultant multiple-objective optimization problem is formulated as
\begin{align}\label{eq74}
\text{Find:} \!\! & \left(d_{\rm opt}^{(d/r)}, d_{\rm opt}^{(r/g)}\right)\!\! \text{ and } \!\!\left(\alpha_{\rm opt},\beta_{\rm opt}\right) \!\! \left\{\!\!\!\!\begin{array}{l} \text{to maximize } \!\! T_r\big(T_{\rm total}\big) , \\
\text{to minimize } t^* , 
 \end{array} \right.
\end{align}
\begin{align}\label{eq75}
\text{subject to:} & \left\{\begin{array}{l} 
 D_{\text{min}}^{(d/r)} \le d_{\rm opt}^{(d/r)} \le D_{\text{max}}^{(d/r)} , \\
 D_{\text{min}}^{(r/g)} \le d_{\rm opt}^{(r/g)} \le D_{\text{max}}^{(r/g)} ,
 \end{array}
 \right.
\end{align} 
where $T_{\rm total}$ is the total time period considered or the predefined working time.

\subsection{$\epsilon$-MOGA assisted Pareto-Optimization}\label{S4-3}

Intuitively, there are no closed-form solutions for the twin-objective optimization problem (\ref{eq74}) and (\ref{eq75}), since the pair of objectives in (\ref{eq74}) should be considered at the same time under the specific constraint of (\ref{eq75}). Hence, we resort to the multi-objective genetic algorithm $\epsilon$-MOGA \cite{reynoso2014controller} in order to acquire the optimal Pareto-front of all solutions of this multi-objective optimization problem. The $\epsilon$-MOGA is an elitist multi-objective evolutionary algorithm based on the concept of $\epsilon$-dominance \cite{reynoso2014controller}, which includes the operations of {\emph{initialization}}, {\emph{Archive}}, {\emph{Variant}}, {\emph{Selection}} and {\emph{Update}} as elaborated on below.

\begin{itemize}
\item [1)] \textbf{Initialization}. At the first generation of $g=1$, where $g$ denotes the generation index, the $\epsilon$-MOGA initializes its population of $P_s$ $4$-element individuals, denoted as $\bm{P}^{(g)}$. Explicitly, the $p_{s}$-th individual is given by
\begin{align}\label{eq76}
\bm{d}_{p_{s}}^{(g)} = \left[d_{p_{s},1}^{(g)} ~ d_{p_{s},2}^{(g)} ~ \alpha_{p_s} ~ \beta_{p_s}\right]^{\rm T} , ~ 1\le p_s \le P_s , 
\end{align}
where $d_{p_{s},1}^{(g)}$ is randomly generated within the range of $[D_{\text{min}}^{(d/r)}, D_{\text{max}}^{(d/r)}]$, and $d_{p_{s},2}^{(g)}$ is randomly generated within the range of $[D_{\text{min}}^{(r/g)}, D_{\text{max}}^{(r/g)}]$, while $\alpha_{p_s}$ is randomly picked from  the range of $(0, ~ 1]$, and $\beta_{p_s}$ in the range of $[0, ~ 1)$.

\vskip 0.1in
\item [2)] \textbf{Archive}. By calculating and comparing the objectives of the throughput $T_r(t_{\rm total})$ and the latency $t^*$ for the population of $\bm{P}^{(g)}$, the $\epsilon$-Pareto-front solution set $\widetilde{\mathbf{R}}$ are selected. Explicitly, the individuals in the $\epsilon$-Pareto-front solution set $\widetilde{\mathbf{R}}$ $\epsilon$-dominate all the other individuals that are not selected for inclusion into $\widetilde{\mathbf{R}}$. An individual $\bm{d}_{p_{s}}^{(g)}$ $\epsilon$-dominates an individual $\bm{d}_{p_{s}^{'}}^{(g)}$ if and only if the objective functions of $\bm{d}_{p_{s}}^{(g)}$ are not worse than the objective functions of $\bm{d}_{p_{s}^{'}}^{(g)}$, and at least one objective function value of $\bm{d}_{p_{s}}^{(g)}$ is better than the same objective function of $\bm{d}_{p_{s}^{'}}^{(g)}$ \cite{reynoso2014controller}.  Furthermore, there is also an elite population archive $\bm{A}^{(g)}$. The individuals in $\widetilde{\mathbf{R}}$ that are not $\epsilon$-dominated by the individuals in $\bm{A}^{(g)}$ will be copied into $\bm{A}^{(g)}$. Note that $\bm{A}^{(1)}$ is initialized as an empty set at the first generation.
\vskip 0.1in
\item [3)] \textbf{Variant}. A new variant is generated by the amalgamation of the `\emph{crossover}' and `\emph{mutation}' operations, which are typically two separate operations in single-objective GA optimization. Specifically, a pair of  individuals, $\bm{r}^{(g,P)}$ and $\bm{r}^{(g,A)}$, are randomly selected, one from the main population $\bm{P}^{(g)}$ and one from the elite population $\bm{A}^{(g)}$, respectively. A randomly generated value $p_{\rm rand}\! \in\! [0, ~ 1]$ is compared to the mutation factor $p_{c/m}$ to decide which operation should be applied to $\bm{r}^{(g,P)}$ and $\bm{r}^{(g,A)}$.

\begin{itemize}
\item [\circled{i}] {\bf{Crossover}}. If $p_{\rm rand} > p_{c/m}$, $\bm{r}^{(g,P)} = \Big[r_{1}^{(g,P)}$ $r_{2}^{(g,P)} ~ r_{3}^{(g,P)} ~ r_{4}^{(g,P)}\Big]^{\rm T}$ and $\bm{r}^{(g,A)} = \Big[r_{1}^{(g,A)}$ $r_{2}^{(g,A)} ~  r_{3}^{(g,A)} ~ r_{4}^{(g,A)}\Big]^{\rm T}$ will cross over using the extended linear recombination, which is formulated as
\begin{align}\label{eq77}
\left\{ \begin{array}{lll}
\widehat{\bm{r}}_1^{(g,G)} &=& \omega \bm{r}^{(g,P)} + (1 - \omega )\bm{r}^{(g,A)}, 
\\
\widehat{\bm{r}}_2^{(g,G)} &=& (1 - \omega )\bm{r}^{(g,P)} + \omega \bm{r}^{(g,A)},
\end{array} \right.
\end{align}
where $\omega$ is a weighting factor of the extended linear recombination \cite{herrero2006robust}.

\item [\circled{ii}] {\bf{Mutation}}. If $p_{\rm rand} \le p_{c/m}$, $\bm{r}^{(g,P)}$ and $\bm{r}^{(g,A)}$ will be mutated using the random mutation associated with the Gaussian distribution \cite{reynoso2014controller}, to yield two new offspring. 
\end{itemize}

The crossover or mutation operations are activated $N_O / 2$ times, which results in a total of $N_O$ new offspring in the auxiliary population $\bm{G}^{(g)}$.

\vskip 0.1in
\item [4)] \textbf{Selection}. The selection operation of multiple-objective optimization is much more complex than that of single-objective optimization. Explicitly, the $\epsilon$-DMOGA calculates the multiple objective function values of the individuals in the auxiliary population $\bm{G}^{(g)}$ and decides which specific individual will be selected into the elite population $\bm{A}^{(g)}$ on the basis of its location in the objective space  \cite{reynoso2014controller}. 

\vskip 0.1in
\item [5)] \textbf{Update}. An individual $\widehat{\bm{r}}_i^{(g,G)}$ from the auxiliary population $\bm{G}^{(g)}$ is compared to an individual $\bm{r}_j^{(g,P)}$ that is randomly selected from the main population $\bm{P}^{(g)}$: if $\widehat{\bm{r}}_i^{(g,G)}$ $\epsilon$-dominates $\bm{r}_j^{(g,P)}$, $\bm{r}_j^{(g,P)}$ is replaced by $\widehat{\bm{r}}_i^{(g,G)}$ in $\bm{P}^{(g)}$. The updating operation is continued until all the individuals in the auxiliary population $\bm{G}^{(g)}$ are compared to an individual randomly selected from the main population $\bm{P}^{(g)}$.

\vskip 0.1in
\item [6)] \textbf{Termination}. The ultimate stopping criterion would be that the Pareto-front solutions of the multiple-objective routing optimization problem have been found. However, we cannot offer any proof of evidence that the Pareto-optimal routing paths have indeed been found. 

In order to have limited and predicable computational complexity, we opt for halting the optimization procedure when the pre-defined maximum affordable number of generations $g_{\max}$ has been exhausted. The individuals from $A^{(g_{\text{max}})}$ then comprise the near-Pareto solutions. Otherwise, we set $g = g + 1$ and go to 2)~\textbf{Archive}.
\end{itemize}

\subsection{Implementation and computational complexity}
In our mobile relaying-assisted drone swarm network architecture, a swarm of drones acting as DCDS for sensing and collecting data using their mounted cameras and/or sensors, whilst a powerful UAV acting as a mobile relaying repeats a round-trip between DCDS and the GS for relaying data from DCDS to GS. Small and micro drones relying on rotor can be used as DCDS due to their low cost and sensing capability, which can be deployed to cover multiple target areas. By contrast, the powerful fixed wing UAVs can be used as RUAV in our  mobile relaying-assisted drone swarm network, which can fly at a much higher speed and have a much longer recharge period as well as  a large-scale antennas. Explicitly, drones acting as DSCS are powered by built-in battery, which typically last 30 minutes. The professional drone DJI Mavic 3 is capable of lasting up to 46 minutes. The powerful RUAV may rely on fixed wing UAV, which uses aerodynamics similar to that of aircraft. It has much longer flight time, namely between 50 and 300 minutes. Nevertheless, the mobile relaying-assisted drone swarm network is indeed energy-critical, which may determine whether the mission can be completed. Wireless power transfer and energy harvesting \cite{huang2019wireless} is a promising technology for powering drones and wireless sensors. However, classical energy harvesting and wireless power transfer is critically dependent on to the charging distance. Oubbati \emph{et al.}~\cite{oubbati2022multiagent} conceived a wireless powering strategy by deploying a set of intelligent flying energy sources operating autonomously. Multiagent deep reinforcement learning was employed for optimizing the energy transfer between the flying energy sources and UAVs. Another potential solution is to use laser-guns for charging  \cite{liu2016charging}. But again, our investigations in this paper do not consider the propulsion power issues, which may be further investigated under the assumption of offloading data to the GS whilst charging the RUAV. Alternatively, a  powerful RUAV can be used as a wireless power station for the DCDS, whilst loading data from DCDS. Pareto optimization of network lifetime, data delivered and delay imposed can be conducted,  while considering the buffer size, battery capacity, loading/offloading points and link adaptation.

In our proposed optimization scheme, we maximize the data delivered in a given time period, whilst minimizing the delay imposed along with considering the working time, communication distance, and buffer size. The computational complexity is bounded by the number of generations $g_{\max}$ and the
population size $P_{s}$. Some additional complexity is imposed
by the crossover and mutation as well as selection operations. Roughly, the computational complexity can be quantified by the number of cost function (CF) evaluations, which is given by $(P_{s} + N_{O})g_{\max}$ CF-evaluation.

The $\epsilon$-MOGA assisted Pareto-Optimization detailed in subsection IV.C can be implemented either online or offline, depending on whether the operating conditions change, such as 
the total distance between the GS and the DCDS, the buffer size of RUAV, the number of DCDSs and the number of antennas activated, as well as the working time (network lifetime). Typically, the buffer size of RUAV, the number of DCDSs and the number of antennas activated will remain unchanged, once the mobile relaying-assisted drone network is established. But the total distance between the GS and the DCDS may be changed, if the DCDS flies to distant areas for sensing and surveillance. The network lifetime is limited by the battery capacity, which typically remains unchanged as well. But some factors may affected the battery capacity, such as the ambient operating temperature, payload, wind and altitude. It would be unsafe to allow a drone operate  until running out battery. Backup drones may be deployed to replace the DCDS following a  specifically designed handover strategy to avoid service interruption. Again, Pareto optimization of the  network lifetime, data delivered and delay imposed as well as wireless powering \cite{oubbati2022multiagent} can be jointly considered in future investigation.

\section{Simulation Experiments}\label{S5}

\begin{figure}[bp!]
\vspace*{-5mm}
\begin{center}
 \includegraphics[width=0.45\textwidth]{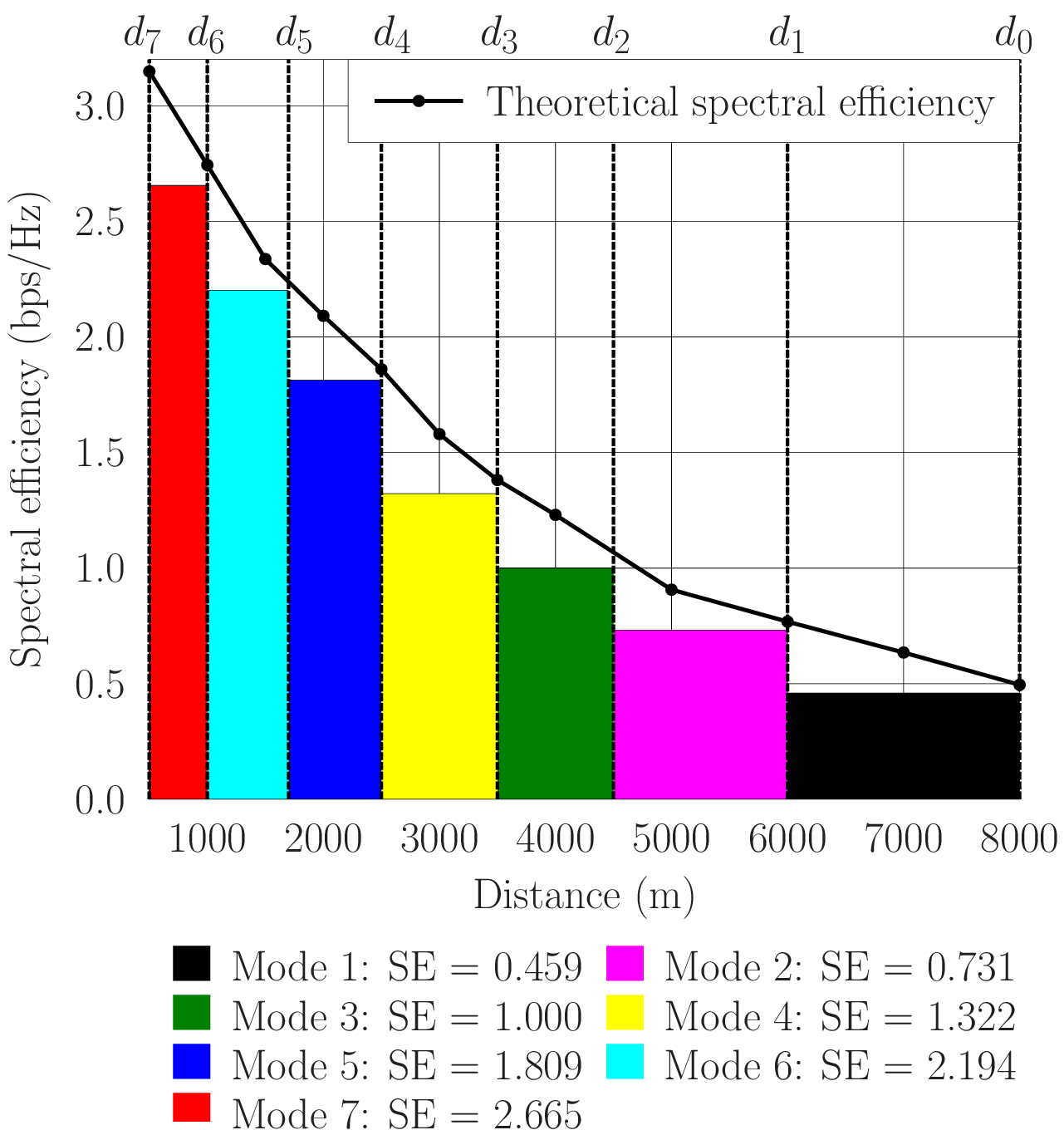}
\end{center}
\vspace*{-2mm}
\caption{An examples of distance-based ACM scheme.}
\label{FIG4}
\vspace*{-1mm}
\end{figure}

In this section, we investigate the achievable performance of our distance-based ACM assisted  RUAV-aided drone swarm communications system consisting of a GS, 4 RUAVs and 32 DCDSs. The GS is serving 4 RUAVs at the same time, whilst each RUAV is capable of simultaneously servicing 8 DCDSs. Specifically, we focus our attention on the achievable performance of the targeted DCDS and RUAV in the presence of realistic interference.  
Traditional aeronautical communications mainly use the very high
frequency band spanning from 118 MHz to 137 MHz, which has been almost fully licensed. Moreover it is impossible to mount large-scale antennas on the UAVs in this frequency range. In order to avoid license restriction whilst providing high-rate aeronautical communications, it is of prime importance to explore
unlicensed frequencies in the millimeter wave (mmWave)
band spanning from 30 GHz to 300 GHz, where the wavelength ranges from 1mm
to 10 mm, resulting in 0.5mm to 5mm TPC antenna spacing.  Hence, the powerful RUAV relying on fixed wing UAVs is capable of carrying a large-scale millimeter wave (mmWave) antenna. Specifically, the wingspan of fixed wing UAV is typically 3 meters, which has enough space to mount hundreds antennas if we use 60\,GHz carrier frequency. Without loss generality, both the GS and the RUAV are equipped with $N_r\! =\! 64$ RAs. Since the size of a drone is much smaller, and it is less powerful in term of load weight and flight duration, the DCDS consists of 8 single-TA drones and hence the number of TAs is $N_t\! =\! 8$. Furthermore, the RUAV will activate $N_t\! =\! 8$ transmit antennas for forwarding the DCDS's messages. The velocity of the RUAV is $50\,{\rm m}/{\rm s}$. The network is allocated  a bandwidth of $B_{\rm total}\! =\! 6$\,MHz at the carrier frequency of 60\,GHz. The transmit power per TA is $P_t\! =\! 78$\,mW. Typically, the UAV channel consists of a LOS path and a cluster of reflected/delayed paths  \cite{matolak2017airground,Bithas2020uav,lian2021anonstationary}. Hence, the drones experience Rician fading, where the Rician factor is set to $K_{\text{Rice}}\! =\! 5$\,dB. We consider a pair of RUAV relaying assisted FANET scenarios based on either static or  mobile relaying. Hence we design two simulation experiments to investigate the achievable performance of the proposed distance-based ACM and RUAV-aided drone swarm. The minimum and maximum distances between the RUAV and the GS/drones are 0.5\,km and 8\,km, respectively. The minimum distance is considered for flight safety. The maximum distance is limited by the maximum communication range beyond which the throughput is zero as illustrated in Fig.~\ref{FIG4} and Table~\ref{Tab2}. To study the impact of the RUAV's buffer size on the achievable performance, both 32\,GB and 64\,GB buffers are considered in our simulations. The default distance-based ACM assisted RUAV-aided drone swarm communications system parameters used for our analysis and simulations are summarised in Table~\ref{Tab1}, whilst the distance-based ACM modes used are  detailed in Table~\ref{Tab2}.

\begin{table*}[btp!]
\caption{Parameters used in validating distance-based ACM assisted RUAV-aided drone swarm communications system}
\begin{center} 
\begin{tabular}{L{4.5cm}|L{6.0cm}|L{5cm}}
\hline\hline
\multirow{16}{4.5cm}{Common system configuration} & The number of RUAV served by a GS & 4\\
&The number of DCDSs served by a RUAV & 8 \\
&The number of RAs for GS & 64 \\
& The number of TAs for GS  & 4\\
& The number of RAs for RUAV & 64 \\
& The number of TAs for RUAV  & 8\\
& The number of TAs for DCDS  & 1\\
& The number of RAs for DCDS  & 1\\
& The velocity of the RUAV & 50 m/s \\
& The carrier frequency $f_{c}$ & 60 GHz \\
& Bandwidth & 6 MHz\\
& The transmit power per TA &  78 mW\\
& The channel used  &  Rice channel\\
& The Rician factor &  5 dB\\
& The buffer size of RUVA &  32 GB and 64 GB\\
&  ACM & As detailed in Table~II \\
\hline
\hline
\multirow{7}{4.5cm}{Scenario I: stationary relay available} & The distance between the
DCDS and the GS $d^{(g/d)}$ &  8.5 km\\
& Maximum end-to-end BE of stationary relay & 1.000 bps/Hz \\
& \multirow{2}{6.0cm}{Location of RUAV for maximum end-to-end BE} & $d_{(*)}^{(r/g)} = 4.25\,\text{km}$  $d_{(*)}^{(r/d)} = 4.25\,\text{km}$ \\
& Minimum end-to-end BE of stationary relay &  0.459 bps/Hz \\
& \multirow{2}{6.0cm}{Location of RUAV for minimum end-to-end BE} & $d_{(*)}^{(r/g)} = 8.0\,\text{km}$  $d_{(*)}^{(r/d)} =0.5\,\text{km}$ \\
& \multirow{2}{6.0cm}{32 GB buffer, multi-objective optimal Solution 1} &  $\Big[d_{\text{32G,opt1}}^{(d/r)} ~ d_{\text{32G,opt1}}^{(r/g)} ~ \alpha_{\text{32G,opt1}} ~ \beta_{\text{32G,opt1}}\Big]\! =\! [3450.5\,\text{m} ~ 632.0\,\text{m} ~ 0.64 ~ 0.11]$ \\
& \multirow{2}{6.0cm}{32 GB buffer, multi-objective optimal Solution 2} &  $\Big[d_{\text{32G,opt2}}^{(d/r)} ~ d_{\text{32G,opt2}}^{(r/g)} ~ \alpha_{\text{32G,opt2}} ~ \beta_{\text{32G,opt2}}\Big]\! =\! [505.5\,\text{m} ~ 576.0\,\text{m} ~ 0.88 ~ 0.12]$ \\
& \multirow{2}{6.0cm}{64 GB buffer, multi-objective optimal Solution 1} &  $\Big[d_{\text{64G,opt1}}^{(d/r)} ~ d_{\text{64G,opt1}}^{(r/g)} ~ \alpha_{\text{64G,opt1}} ~ \beta_{\text{64G,opt1}}\Big]\! = \! [3496.9\,\text{m} ~ 586.2\,\text{m} ~ 0.50 ~ 0.07]$ \\
& \multirow{2}{6.0cm}{64 GB buffer, multi-objective optimal Solution 2} & $\Big[d_{\text{64G,opt2}}^{(d/r)} ~ d_{\text{64G,opt2}}^{(r/g)} ~ \alpha_{\text{64G,opt2}} ~ \beta_{\text{64G,opt2}}\Big]\! =\! [777.2\,\text{m} ~ 769.3\,\text{m} ~ 0.67 ~ 0.05]$ \\
\hline
\hline
\multirow{19}{4.5cm}{Scenario II: stationary relay unavailable} & The distance between the
DCDS and the GS $d^{(g/d)}$ &  25 km\\
& Near-GS point as illustrated in Fig.~1 (b) & $d^{(r/g)} = 0.5\,\text{km}$ \\
& Near-DCDS point as illustrated in Fig.~1 (b) & $d^{(r/g)} = 24.5\,\text{km}$ \\
& 32 GB buffer, benchmark Solution-1 &  $\Big[d_{\text{32G,b1}}^{(d/r)} ~ d_{\text{32G,b1}}^{(r/g)} ~ \alpha_{\text{32G,b1}} ~ \beta_{\text{32G,b1}}\Big]\! =\! [500.0\,\text{m} ~ 500.0\,\text{m} ~ 1.0 ~ 0]$ \\
& 32 GB buffer, benchmark Solution-2 &  $\Big[d_{\text{32G,b2}}^{(d/r)} ~ d_{\text{32G,ob2}}^{(r/g)} ~ \alpha_{\text{32G,b2}} ~ \beta_{\text{32G,b2}}\Big]\! =\! [7999.9\,\text{m} ~ 7999.9\,\text{m} ~ 1.0, 0]$ \\
& \multirow{2}{6.0cm}{32 GB buffer, multi-objective optimal Solution 1} &  $\Big[d_{\text{32G,opt1}}^{(d/r)} ~ d_{\text{32G,opt1}}^{(r/g)} ~ \alpha_{\text{32G,opt1}} ~ \beta_{\text{32G,opt1}}\Big]\! =\! [953.0\,\text{m} ~ 510.2\,\text{m} ~ 0.50 ~ 0.13]$ \\
& \multirow{2}{6.0cm}{32 GB buffer, multi-objective optimal Solution 2} &  $\Big[d_{\text{32G,opt2}}^{(d/r)} ~ d_{\text{32G,opt2}}^{(r/g)} ~ \alpha_{\text{32G,opt2}} ~ \beta_{\text{32G,opt2}}\Big]\! =\! [779.6\,\text{m} ~ 547.2\,\text{m} ~ 0.60 ~ 0.26]$ \\
& 64 GB buffer, benchmark Solution-1 &  $\Big[d_{\text{32G,b1}}^{(d/r)} ~ d_{\text{32G,b1}}^{(r/g)} ~ \alpha_{\text{32G,b1}} ~ \beta_{\text{32G,b1}}\Big]\! =\! [500.0\,\text{m} ~ 500.0\,\text{m} ~ 1.0 ~ 0]$ \\
& 64 GB buffer, benchmark Solution-2 &  $\Big[d_{\text{32G,b2}}^{(d/r)} ~ d_{\text{32G,ob2}}^{(r/g)} ~ \alpha_{\text{32G,b2}} ~ \beta_{\text{32G,b2}}\Big]\! =\! [7999.9\,\text{m} ~ 7999.9\,\text{m} ~ 1.0, 0]$ \\
& \multirow{2}{6.0cm}{64 GB buffer, multi-objective optimal Solution 1} &  $\Big[d_{\text{64G,opt1}}^{(d/r)} ~ d_{\text{64G,opt1}}^{(r/g)} ~ \alpha_{\text{64G,opt1}} ~ \beta_{\text{64G,opt1}}\Big]\! = \! [829.5\,\text{m} ~ 3459.3\,\text{m} ~ 0.50 ~ 0]$ \\
& \multirow{2}{6.0cm}{64 GB buffer, multi-objective optimal Solution 2} & $\Big[d_{\text{64G,opt2}}^{(d/r)} ~ d_{\text{64G,opt2}}^{(r/g)} ~ \alpha_{\text{64G,opt2}} ~ \beta_{\text{64G,opt2}}\Big]\! =\! [839.3\,\text{m} ~ 523.2\,\text{m} ~ 0.85 ~ 0.08]$ \\
\hline
\end{tabular}
\end{center}
\label{Tab1}
\end{table*}

\subsection{Distance-based ACM}\label{S5.1}

The theoretically achievable rate per TA as a function of distance is indicated by the solid curve marked by dots in Fig.~\ref{FIG3}. By designing the eight distance thresholds $d_q$ for $0 \le q \le 7$ to ensure that the rate of mode $q$ is lower than the theoretically achievable rate in the distance range $[d_q , ~ d_{q-1}]$, we obtain the corresponding six desired distance thresholds for this ACM, which are indicated in Fig.~\ref{FIG3}. Note that $d_{0}$ and $d_{7}$ represent the near-GS point and the near-DCDS point, respectively, as illustrated in Fig.~1 (b). The seven ACM modes used and the associated modulations schemes as well as coding rates, are shown in Table~\ref{Tab2}. 

\begin{table*}[tp!]
\vspace*{-2mm}
\caption{Distance-based adaptive coding and modulation scheme for aeronautical communications.}
\vspace*{-4mm}
\begin{center}
\resizebox{\textwidth}{!}{
\begin{tabular*}{17cm}{@{\extracolsep{\fill}}cccccc}
\toprule
 Mode $q$ & Mode color & Modulation & Code rate & Spectral efficiency\,(bps/Hz) & Switching threshold $d_q$\,(km) \\ \toprule
 0 & /        & /      & 0     & 0         & 8.0      \\ \midrule
 1 & Black    & BPSK   & 0.488 & 0.459     & 6.0      \\ \midrule
 2 & Magenta  & BPSK   & 0.780 & 0.731     & 4.5      \\ \midrule
 3 & Green    & QPSK   & 0.533 & 1.000     & 3.5      \\ \midrule
 4 & Yellow   & QPSK   & 0.706 & 1.322     & 2.5      \\ \midrule
 5 & Blue     & 8-QAM  & 0.642 & 1.809     & 1.7      \\ \midrule
 6 & Cyan     & 8-QAM  & 0.780 & 2.194     & 1.0      \\ \midrule
 7 & Red      & 16-QAM & 0.708 & 2.665     & 0.5      \\ \bottomrule
\end{tabular*}
}
\end{center}
\label{Tab2}
\vspace*{-4mm}
\end{table*}

\subsection{Scenario I: stationary relay is available}\label{S5.2}

In \emph{Scenario~I}, both the RUAV-to-DCDS link and the RUAV-to-GS link exist at the same time. As shown in Fig.~\ref{FIG2}, there are multiple cases of \emph{Scenario~I} depending on the distance between the DCDS and the GS. As an example of our investigation for \emph{Scenario~I}, the distance between the DCDS and the GS is 8.5\,km and the RUAV hovers between them,  corresponding to Case~3 of Fig.~\ref{FIG2}. But this investigation is equally applicable to the other cases upon simply changing the related parameter setting. Recalling the analysis of Subsection~\ref{S4-1}, the maximum end-to-end throughput of the RUAV acting as a static relay can be achieved when the RUAV's distance to the GS $d^{(r/g)}$ is in the range of $[4.0\,\text{km},~  4.5\,\text{km}]$. We select the middle point between the DCDS and the GS as the location where the RUAV hovers, i.e., we have $d_{(*)}^{(r/g)}\! =\! 4.25\,\text{km}$ and $d_{(*)}^{(r/d)}\! =\!4.25\,\text{km}$. The achievable maximum end-to-end  throughput is 1.000\,bps/Hz per TA, whilst the total throughput is 8.000\,bps/Hz of all the $N_{t}\! =\! 8$ TAs. Again, as illustrated in Fig.~2 (b), if the RUAV hovers at the near-DCDS point $d_{(*)}^{(r/g)}\! =\! 8.0$\,km  or the near GS point $d_{(*)}^{(r/g)}\! =\! 0.5$\,km, it can only achieve a minimum end-to-end throughput of 0.459\,bps/Hz per TA, while the maximum throughput is 3.672\,bps/Hz for all the $N_{t}\! =\! 8$ TAs.

Naturally, the RUAV is also capable of acting as a mobile relay. We also want to know whether upon acting as a mobile relay it can provide a higher end-to-end throughput without imposing extra delay. When the RUAV acts as a mobile relay, it circles back and forth between the near-DCDS point $d_{\text{opt}}^{(d/r)}$ and the near-GS point at $d_{\text{opt}}^{(r/g)}$\,km. Explicitly, the RUAV hovers at the near-DCDS point $d_{\text{opt}}^{(r/d)}$\,km to receive the data collected by the eight DCDSs at a maximum potential throughput. When the data gleaned fills at a certain percentage $\alpha_{\text{opt}}$ of its buffer, it will fly to the near-GS point at  $d_{\text{opt}}^{(r/g)}$ to offload the data to the GS. Note that there is also some additional  end-to-end data transmission at the throughput of $\min\{R_{\text{total}}^{r/g}, R_{\text{total}}^{d/r}\}$, since both the RUAV-to-DCDS link and the RUAV-to-GS link exist at the same time. The Pareto optimal multiple-objective solutions $\Big[d_{\text{opt}}^{(d/r)} ~ d_{\text{opt}}^{(r/g)} ~ \alpha_{\text{opt}} ~ \beta_{\text{opt}}\Big]$ of both the near-DCDS point, of the near-GS point, as well as the maximum factor of caching data, and the minimum factor of offloading data are also affected by the buffer size. 

\begin{figure*}[tbp!]
\vspace{-4mm}
\begin{center}
 \subfigure[Buffer size: 32GB]{
  \includegraphics[width=0.47\textwidth]{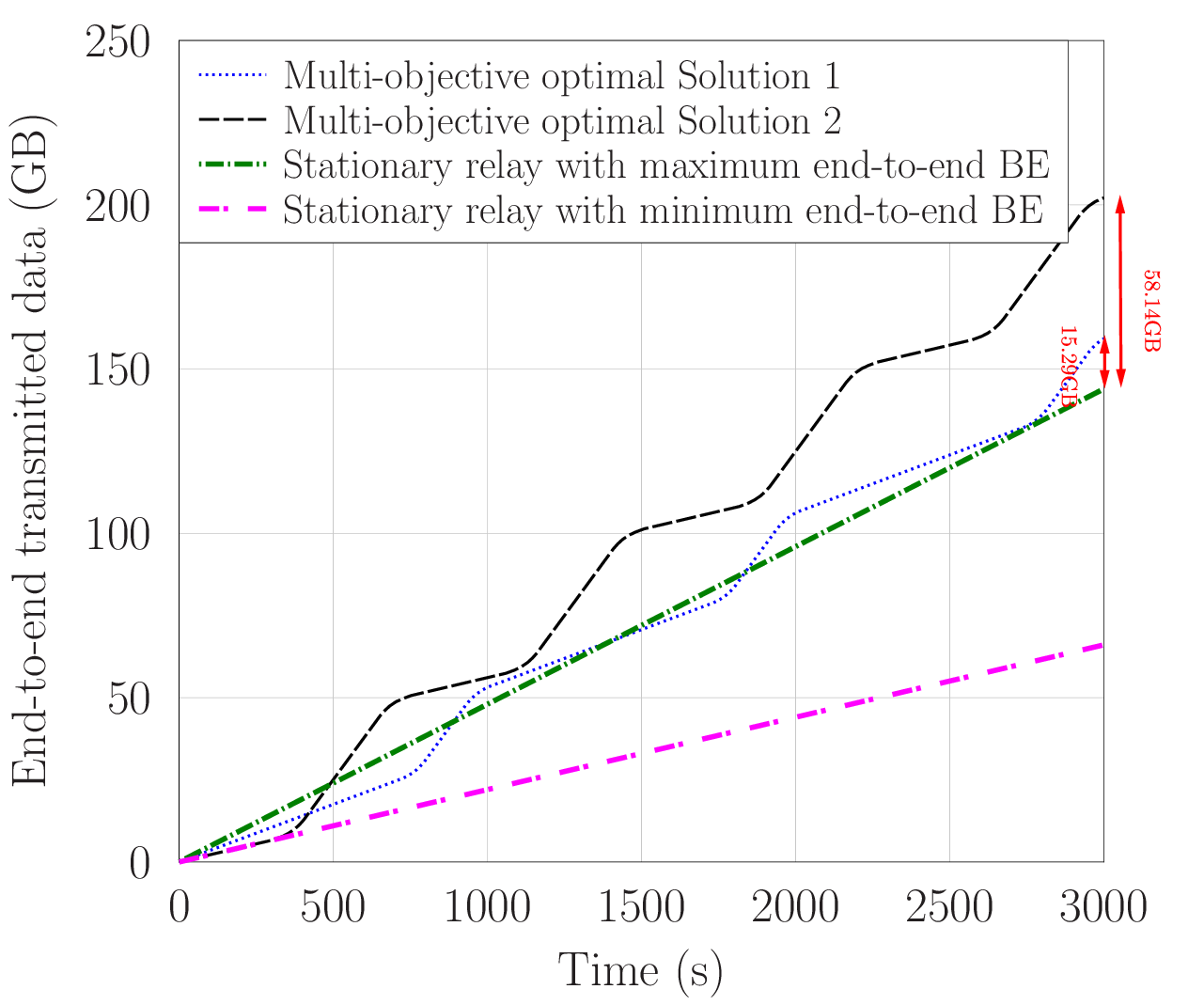} 
  \label{fig5a}
 }%
 \subfigure[Buffer size: 64GB]{
  \includegraphics[width=0.47\textwidth]{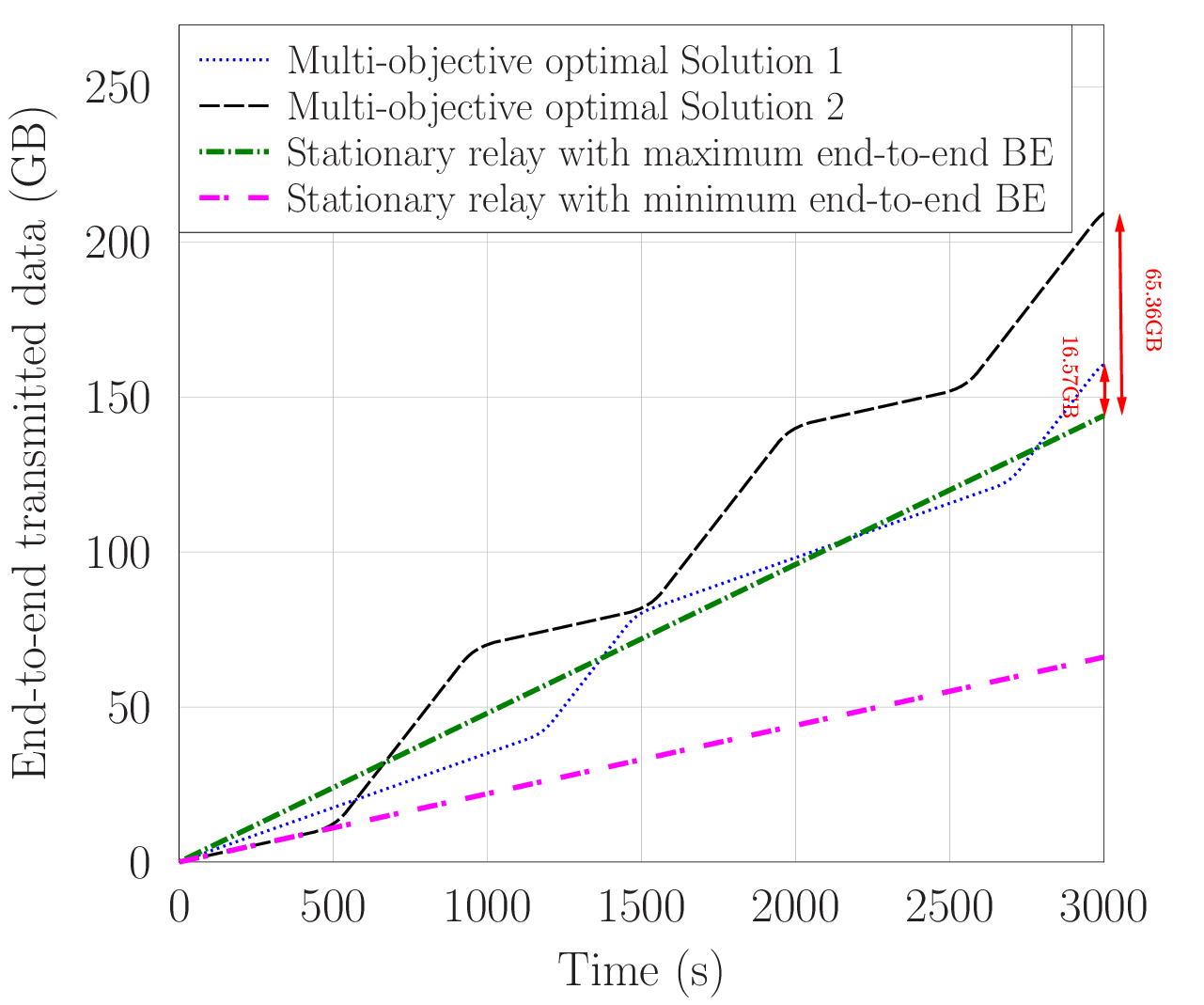}
  \label{fig5b}
 }
\end{center}
\vspace{-2mm}
\caption{The total data transmitted in \emph{Scenario~I}.}
\label{fig5}
\end{figure*}

Firstly in Fig.~\ref{fig5}, we investigate the total amount of data transmitted given the time period of 50 minutes.  Explicitly, Fig.~\ref{fig5a} depicts the performance achieved when the RUAV buffer size is 32\,GB, whilst Fig.~\ref{fig5b} depicts the performance achieved when the RUAV buffer size is 64\,GB. Naturally, the buffer size has no impact on the stationary relay. The stationary relay associated with the minimum end-to-end BE delivers the minimum total amount of data to the GS, where again BE represents bandwidth efficiency. By contrast, the stationary relay having the maximum end-to-end BE is capable of delivering about 124.3\,GB more data than the stationary relay having the minimum end-to-end BE.

\begin{figure*}[tbp!]
\vspace{-4mm}
\begin{center}
 \subfigure[Buffer size: 32GB]{
  \includegraphics[width=0.47\textwidth]{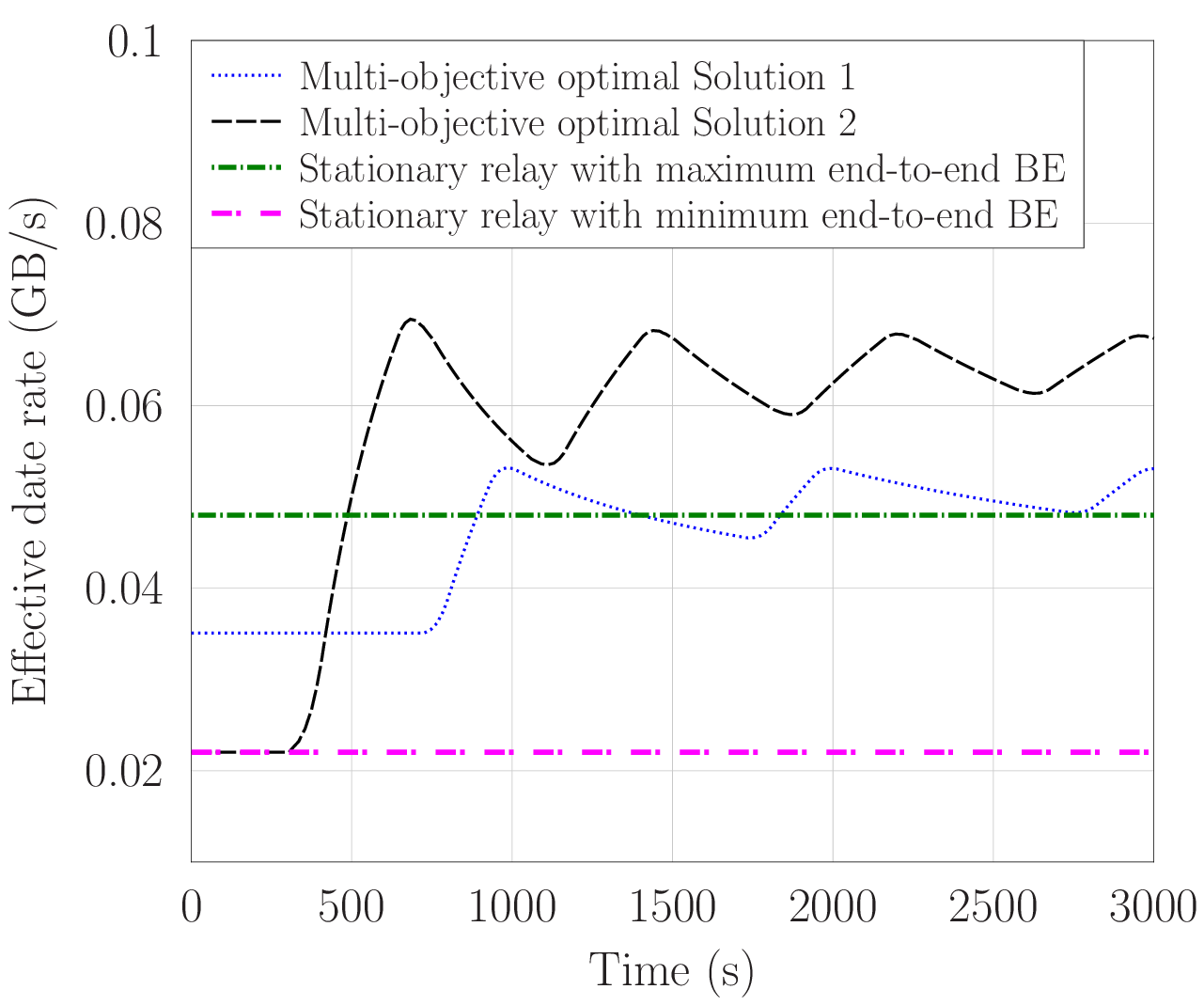} 
  \label{fig6a}
 }%
 \subfigure[Buffer size: 64GB]{
  \includegraphics[width=0.47\textwidth]{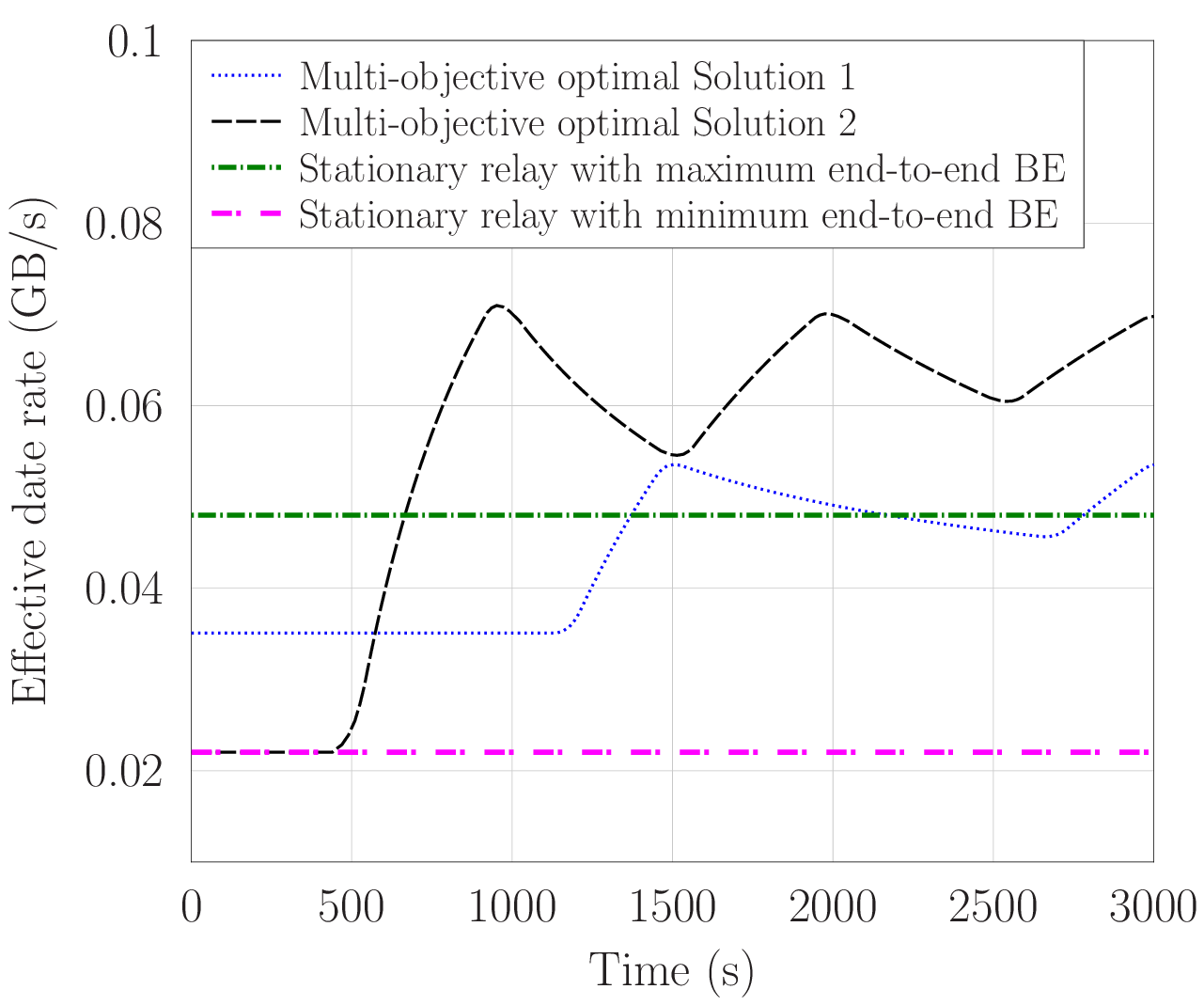}
  \label{fig6b}
 }
\end{center}
\vspace{-2mm}
\caption{The effective data rate as a function of time in \emph{Scenario~I}.}
\label{fig6}
\vspace{-3mm}
\end{figure*}

When we exploit the mobility of the RUAV as a mobile relay, there are two Pareto optimal solutions for the RUAV having 32\,GB buffer, which are $\Big[d_{\text{32G,opt1}}^{(d/r)} ~ d_{\text{32G,opt1}}^{(r/g)} ~ \alpha_{\text{32G,opt1}}$ $\beta_{\text{32G,opt1}}\Big]\! =\! [3450.5\,\text{m} ~ 632.0\,\text{m} ~ 0.64~ 0.11]$ and $\Big[d_{\text{32G,opt2}}^{(d/r)} ~ d_{\text{32G,opt2}}^{(r/g)} ~ \alpha_{\text{32G,opt2}} ~ \beta_{\text{32G,opt2}}\Big]\! =\! [505.5\,\text{m}$ $576.0\,\text{m} ~ 0.88 ~ 0.12]$, respectively. The multiple-objective Pareto optimal Solution 2 is capable of delivering 58.14\,GB more data to the GS than the stationary relay having the maximum end-to-end BE. In normalized terms, it delivered 40.38\% more data. The multiple-objective Pareto  optimal Solution 1 delivers 15.29\,GB more data than the stationary relay having the maximum end-to-end BE. But it imposes a shorter delay than the multiple-objective Pareto  optimal Solution 2. The delay imposed is defined as the time of the effective end-to-end BE becomes higher than that of the stationary relay having the minimum end-to-end BE (see Eq.~(\ref{eq73})), which can be observed in Fig.~\ref{fig6}. 

When the buffer size of the RUAV is 64\,GB, there are also two Pareto-front optimal solutions, which are given by $\Big[d_{\text{64G,opt1}}^{(d/r)} ~ d_{\text{64G,opt1}}^{(r/g)} ~ \alpha_{\text{64G,opt1}} ~ \beta_{\text{64G,opt1}}\Big]\! =\! [3496.9\,\text{m} ~ 586.2\,\text{m} ~ 0.50 ~ 0.07]$ and $\Big[d_{\text{64G,opt2}}^{(d/r)} ~ d_{\text{64G,opt2}}^{(r/g)} ~ \alpha_{\text{64G,opt2}} ~ \beta_{\text{64G,opt2}}\Big]\! =\! [777.2\,\text{m} ~ 769.3\,\text{m} ~ 0.67 ~ 0.05]$, respectively. It can be seen from Fig.~\ref{fig5b} that the Pareto optimal Solution 2 is capable of delivering 65.36\,GB more data in 50 minutes than the RUAV acting as a stationary relay having the maximum end-to-end BE. Explicitly, it delivered 45.38\% more data. Similarly, the multiple-objective Pareto optimal Solution 1 delivers 16.57\,GB more data than the stationary relay having the maximum end-to-end BE, whilst imposing a shorter delay than the multiple-objective Pareto optimal Solution 2.

In Fig.~\ref{fig6}, we investigate the effective end-to-end average data rates $R_{e}(t)$ as the functions of time. Intuitively, both the stationary relay having the maximum end-to-end rate and the stationary relay having the minimum end-to-end rate have constant effective end-to-end average data rates, which are given by $4.80 \times 10^{-2}$\,GB/s and $2.20 \times 10^{-2}$\,GB/s, respectively. However, the effective end-to-end average data rate, defined in (\ref{eq72}) fluctuates when the RUAV acts as mobile relay, which is caused by switching ACM modes in line with the communication distance in order to maximally exploit the link capacity. Observe from Fig.~\ref{fig6a} for the buffer size of 32\,GB that the effective end-to-end average data rate of the multiple-objective Pareto optimal Solution 2 is always higher than that of the stationary relay having the maximum end-to-end rate when the time passes 500\,s. Furthermore, it is higher than the effective end-to-end average data rate of the multiple-objective Pareto optimal Solution 1 for $t \ge 400$\,s. If we consider the rate of the stationary relay having the minimum end-to-end rate as the required minimum effective end-to-end average data rate $R_e^*$, the delay as defined in Eq.\,(\ref{eq73}) becomes $t^*\! = \! 0$\,s for the multiple-objective Pareto optimal Solution 1. By contrast, the delay imposed by the multiple-objective Pareto optimal Solution 2 is $t^*\! =\! 300$\,s. Similar trends can be observed in Fig.~\ref{fig6b} for the buffer size of 64\,GB.

\begin{figure*}[tbp!]
\vspace{-4mm}
\begin{center}
 \subfigure[Buffer size: 32GB]{
  \includegraphics[width=0.47\textwidth]{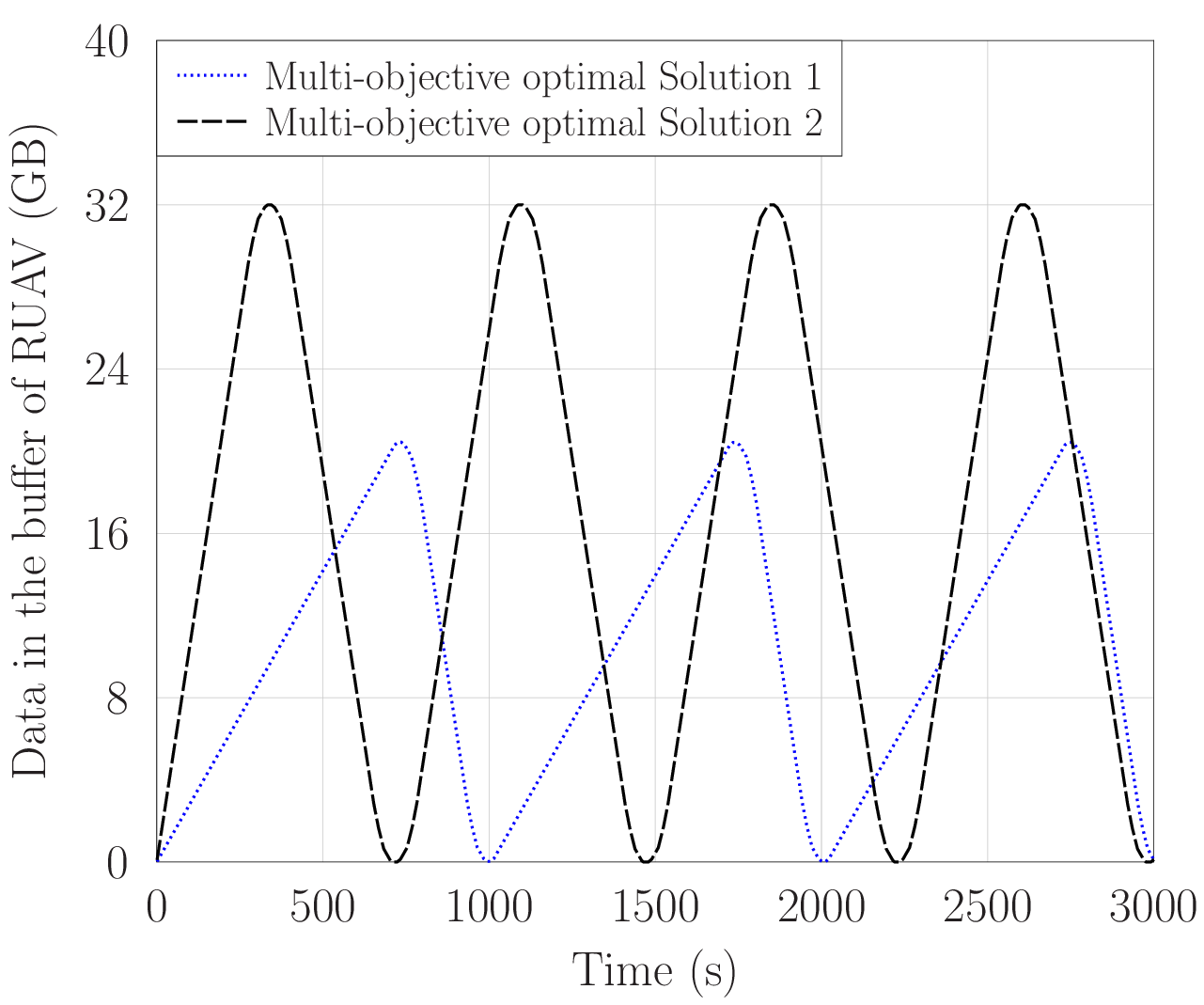} 
  \label{fig7a}
 }%
 \subfigure[Buffer size: 64GB]{
  \includegraphics[width=0.47\textwidth]{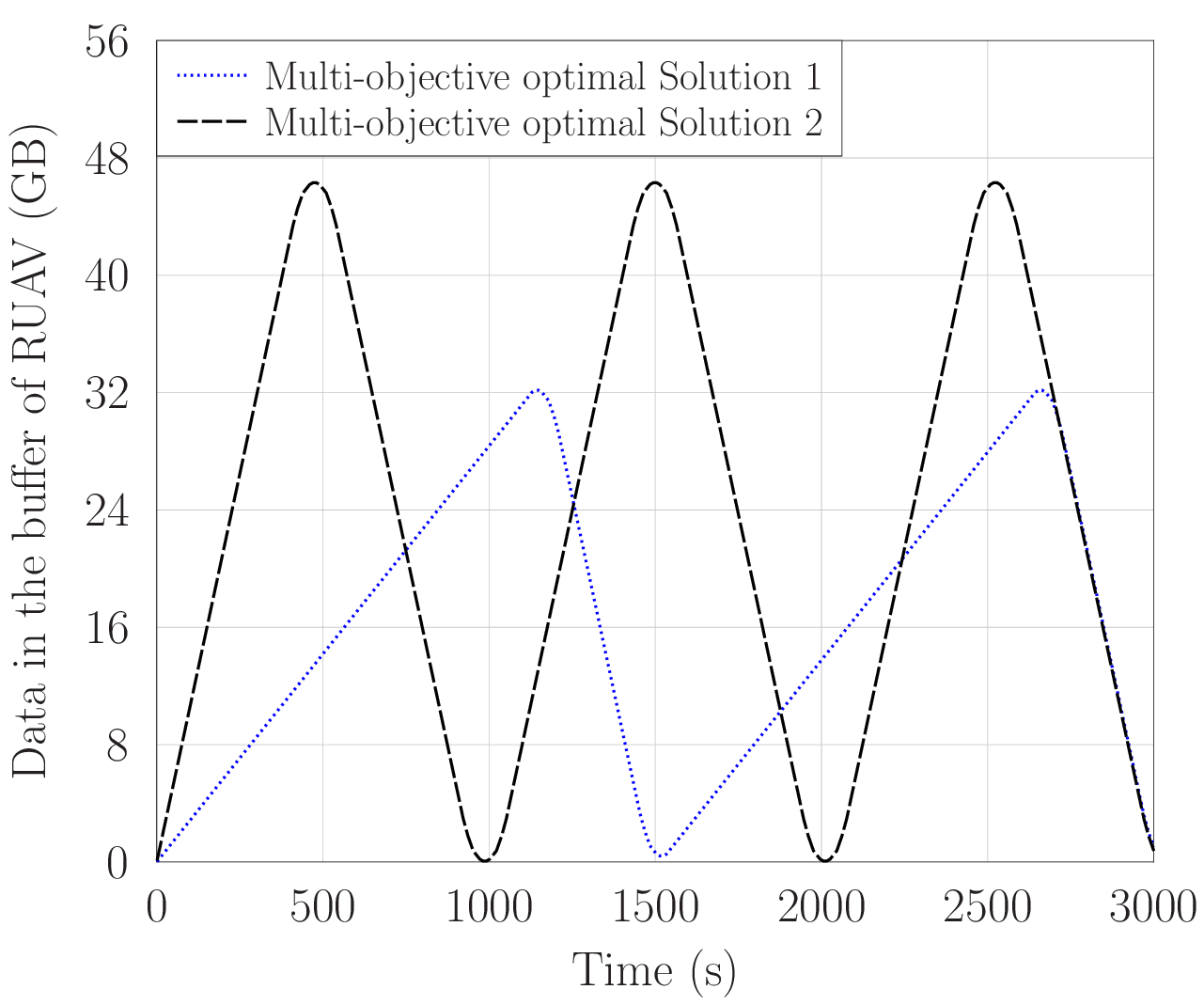}
  \label{fig7b}
 }
\end{center}
\vspace{-2mm}
\caption{Data cached in the RUAV buffer as a function of time in \emph{Scenario~I}.}
\label{fig7}
\vspace{-3mm}
\end{figure*}

The amount of data cached in the buffer of the RUAV versus time can be observed from Fig.~\ref{fig7}. When the RUAV acts as a stationary relay, no data is cached in the buffer. Hence we only plot the data cached in the buffer when the RUAV acts as mobile relay. It can be seen from both Fig.~\ref{fig7a} and Fig.~\ref{fig7b} that the multiple-objective Pareto optimal Solution 2 fully exploits the capacity of the buffer and delivers the maximum data from the DCDS to the GS, but it imposes a longer delay, when aiming for reaching the required minimum effective end-to-end average data rate $R_e^*$, as seen in Fig.~\ref{fig6}. By contrast, the multiple-objective Pareto optimal Solution 1 does not fully exploit the capacity of the buffer and delivers less data from the DCDS to the GS than the multiple-objective Pareto optimal Solution 2, but it imposes a shorter delay.

\subsection{Scenario II: stationary relay is unavailable}\label{S5.3}

In \emph{Scenario~II}, even the minimum-rate most robust communication link may only exist either for the RUAV-to-DCDS or for the RUAV-to-GS.  Explicitly, when the distance between the DCDS and the GS is longer than 16\,000\,m, it is impossible to establish both the RUAV-to-DCDS link and the RUAV-to-GS link at the same time. As a specific example, we set the distance between the DCDS and the GS to 25\,000\,m. The minimum and maximum distances between the RUAV and the GS/drones are 500\,m and 24\,500\,m, respectively. Recall from Fig.~\ref{FIG3} that the maximum communication distance is 8\,000\,m, which means that when the distance between the RUAV and GS/drones exceeds 8\,000\,m, there is no communication link. 

\begin{figure*}[tbp!]
\vspace{-4mm}
\begin{center}
 \subfigure[Buffer size: 32GB]{
  \includegraphics[width=0.47\textwidth]{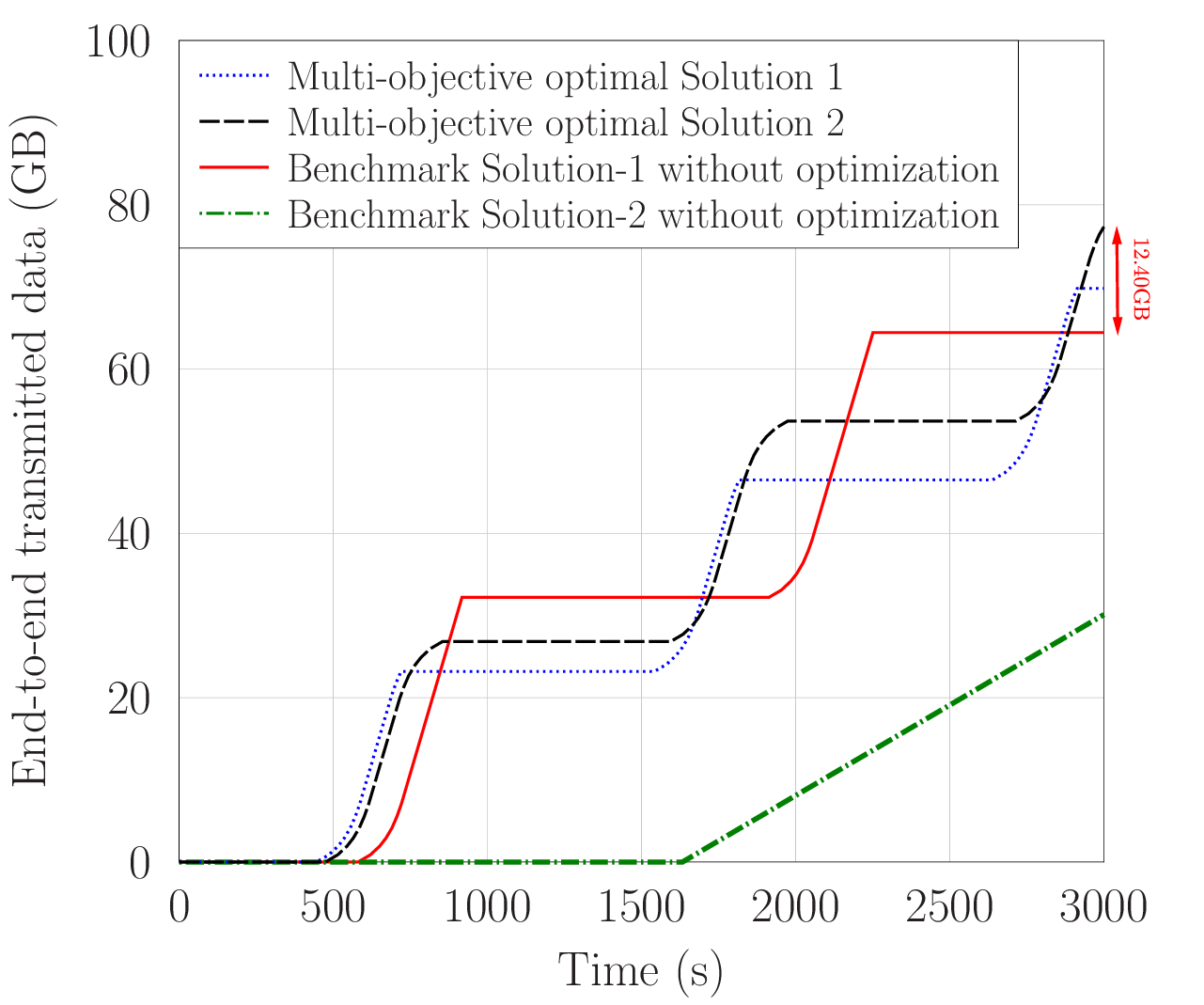} 
  \label{fig8a}
 }%
 \subfigure[Buffer size: 64GB]{
  \includegraphics[width=0.47\textwidth]{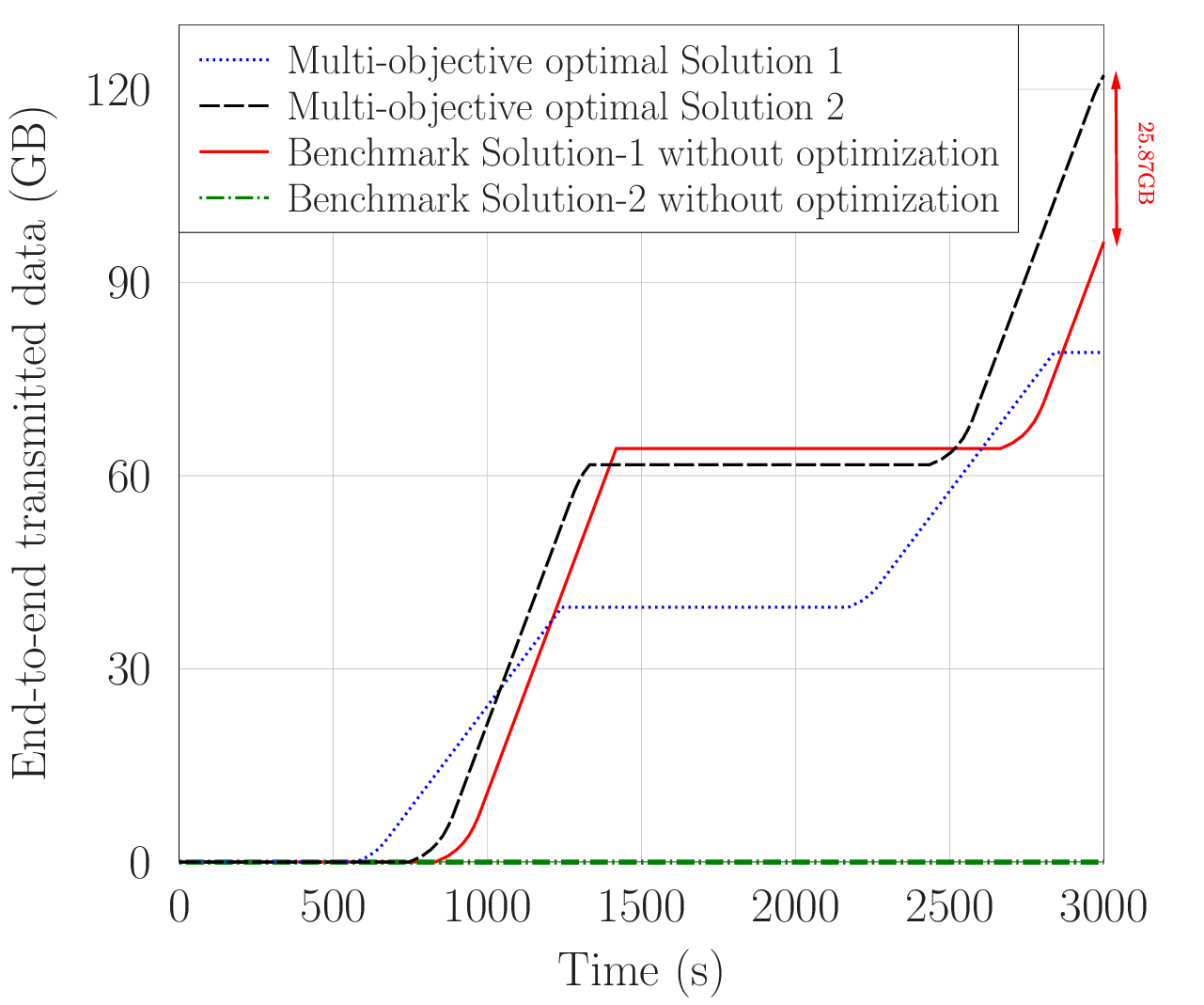}
  \label{fig8b}
 }
\end{center}
\vspace{-2mm}
\caption{The total data transmitted in \emph{Scenario~II}.}
\label{fig8}
\vspace*{-3mm}
\end{figure*}

When the buffer size of the RUAV is 32\,GB, there are 29 Pareto optimal solutions. Here we only characterize the solution having the minimum delay and the solution having the maximum data delivered, which are $\Big[d_{\text{32G,opt1}}^{(d/r)} ~ d_{\text{32G,opt1}}^{(r/g)} ~ \alpha_{\text{32G,opt1}} ~ \beta_{\text{32G,opt1}}\Big]\! =\! [953.0\,\text{m} ~ 510.2\,\text{m} ~ 0.50 ~ 0.13]$ and $\Big[d_{\text{32G,opt2}}^{(d/r)} ~ d_{\text{32G,opt2}}^{(r/g)} ~ \alpha_{\text{32G,opt2}} ~ \beta_{\text{32G,opt2}}\Big]\! =\! [779.6\,\text{m} ~ 547.2\,\text{m} ~ 0.60 ~ 0.26]$, respectively. As a comparison, we also include two solutions without any optimization as our benchmarks, which are the nearest-loading-point and nearest-offloading-point solutions as well as the farthest-loading-point and farthest-offloading-point solution. Explicitly, they are given by $\Big[d_{\text{32G,b1}}^{(d/r)} ~ d_{\text{32G,b1}}^{(r/g)} ~ \alpha_{\text{32G,b1}} ~ \beta_{\text{32G,b1}}\Big]\! =\! [500.0\,\text{m} ~ 500.0\,\text{m} ~ 1.0 ~ 0]$ and $\Big[d_{\text{32G,b2}}^{(d/r)} ~ d_{\text{32G,ob2}}^{(r/g)} ~ \alpha_{\text{32G,b2}} ~ \beta_{\text{32G,b2}}\Big]\! =\! [7999.9\,\text{m} ~ 7999.9\,\text{m} ~ 1.0, 0]$, respectively. 

When the buffer size of the RUAV is 64\,GB, there are 25 Pareto optimal solutions. We characterize the solution having the minimum delay and the solution having the maximum data delivered, given by $\Big[d_{\text{64G,opt1}}^{(d/r)} ~ d_{\text{64G,opt1}}^{(r/g)} ~ \alpha_{\text{64G,opt1}} ~ \beta_{\text{64G,opt1}}\Big]\! = \! [829.5\,\text{m} ~ 3459.3\,\text{m} ~ 0.50 ~ 0]$ and $\Big[d_{\text{64G,opt2}}^{(d/r)} ~ d_{\text{64G,opt2}}^{(r/g)} ~ \alpha_{\text{64G,opt2}} ~ \beta_{\text{64G,opt2}}\Big]\! =\! [839.3\,\text{m} ~ 523.2\,\text{m} ~ 0.85 ~ 0.08]$, respectively. In this case, the nearest-loading-point and nearest-offloading-point solution as well as the farthest-loading-point and farthest-offloading-point solution are given by $\Big[d_{\text{64G,b1}}^{(d/r)} ~ d_{\text{64G,b1}}^{(r/g)} ~ \alpha_{\text{64G,b1}} ~ \beta_{\text{64G,b1}}\Big]\! =\! [500.0\,\text{m} ~ 500.0\,\text{m} ~ 1.0 ~ 0]$ and $\Big[d_{\text{64G,b2}}^{(d/r)} ~ d_{\text{64G,ob2}}^{(r/g)} ~ \alpha_{\text{64G,b2}} ~ \beta_{\text{64G,b2}}\Big]\! =\! [7999.9\,\text{m} ~ 7999.9\,\text{m} ~ 1.0, 0]$, respectively, which are identical to the 32\,GB buffer scenario.

The amount of total data transmitted from the DCDS to the GS is investigated in Fig.~\ref{fig8}. Observe from Fig.~\ref{fig8a} that both the multiple-objective Pareto optimal solutions are capable of delivering more data than the pair of benchmark Solutions, when the buffer size is 32\,GB. The multiple-objective Pareto optimal Solution 2 delivers the most data from the DCDS to the GS, regardless of the buffer size. Explicitly, it delivers 12.4\,GB more data from the DCDS to the GS than the benchmark solution 1 when the buffer size is 32G, and 25.87\,GB more data  than the benchmark Solution 1, when the buffer size is 64\,GB. In other words, our solution is capable of delivering 19.24\% and 26.86\% extra data compared to the benchmark Solution 1 when the buffer sizes are 32\,GB and 64\,GB, respectively. The benchmark Solution 2 delivers the minimum data from the DCDS to the GS. In particular, it delivers no data to the GS in the period of 3000\,s for the buffer size of 64\,GB, because the RUAV has just completed its data loading action at the near-DCDS loading point and it is heading to the GS, but it has not yet reached  the communication range of the GS.

\begin{figure*}[tbp!]
\vspace{-4mm}
\begin{center}
 \subfigure[Buffer size: 32GB]{
  \includegraphics[width=0.47\textwidth]{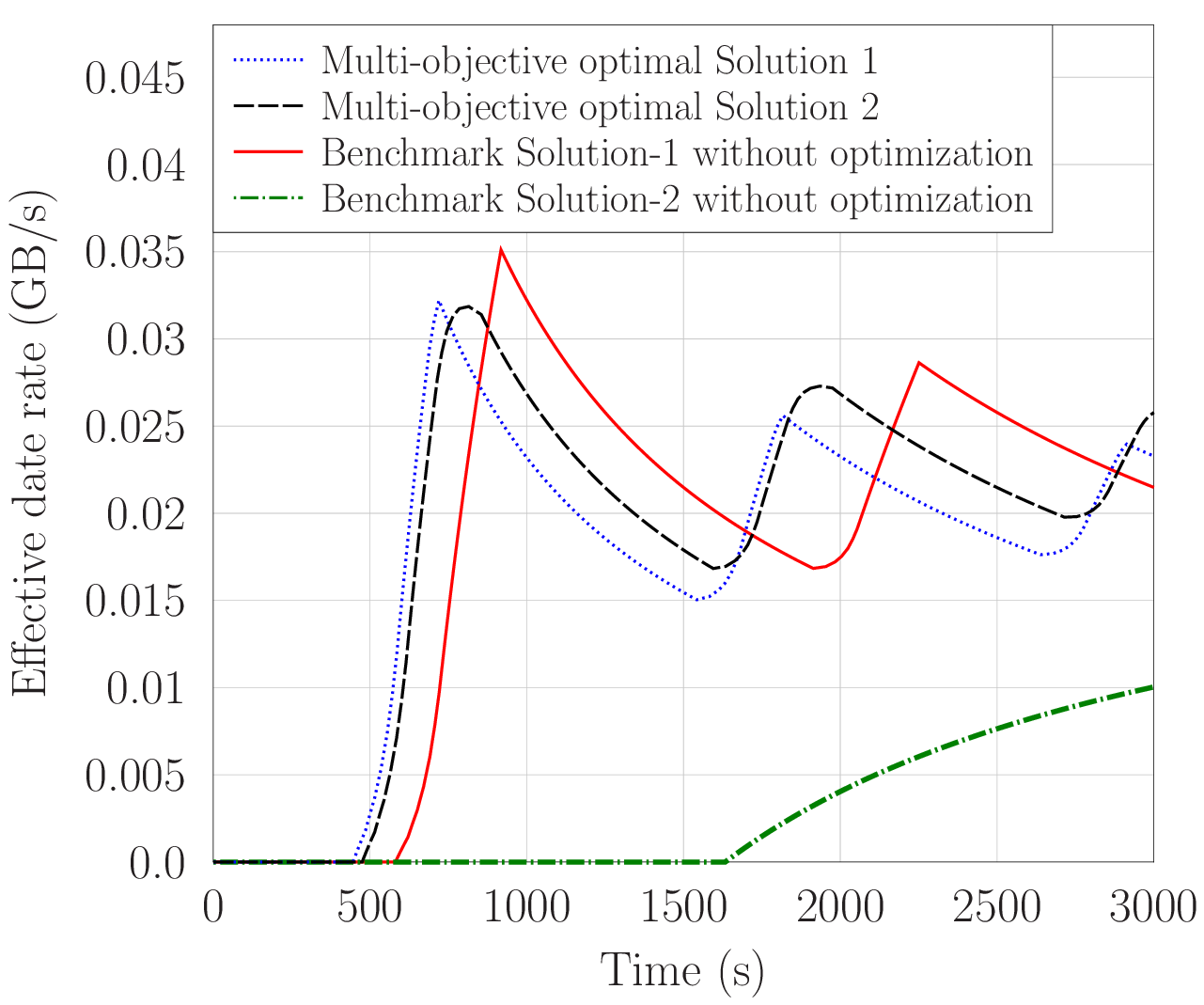} 
  \label{fig9a}
 }%
 \subfigure[Buffer size: 64GB]{
  \includegraphics[width=0.47\textwidth]{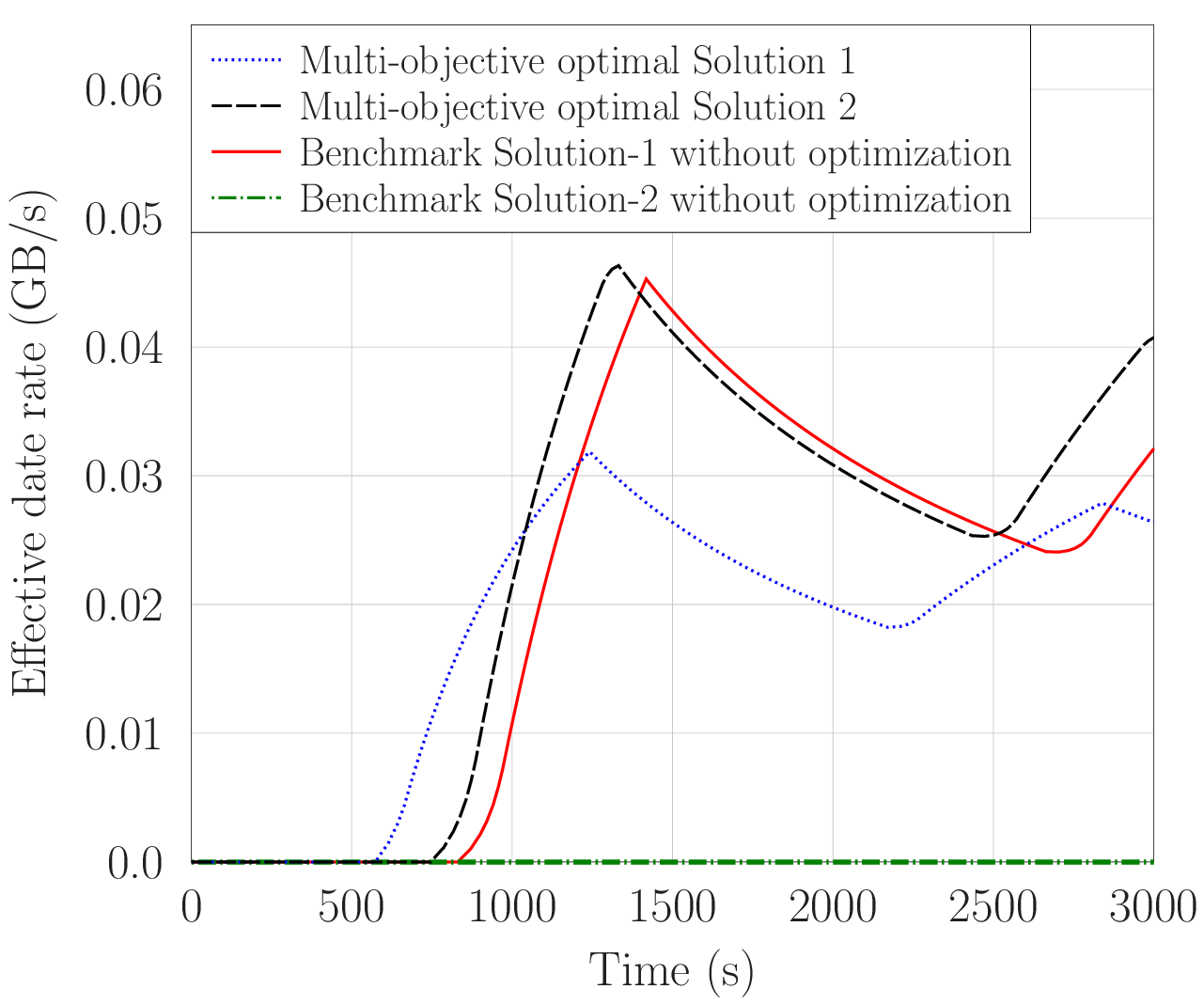}
  \label{fig9b}
 }
\end{center}
\vspace{-2mm}
\caption{The effective data rate as a function of time in \emph{Scenario~II}.}
\label{fig9}
\vspace{-3mm}
\end{figure*}

\begin{figure*}[tbp!]
\vspace{-4mm}
\begin{center}
 \subfigure[Buffer size: 32G]{
  \includegraphics[width=0.47\textwidth]{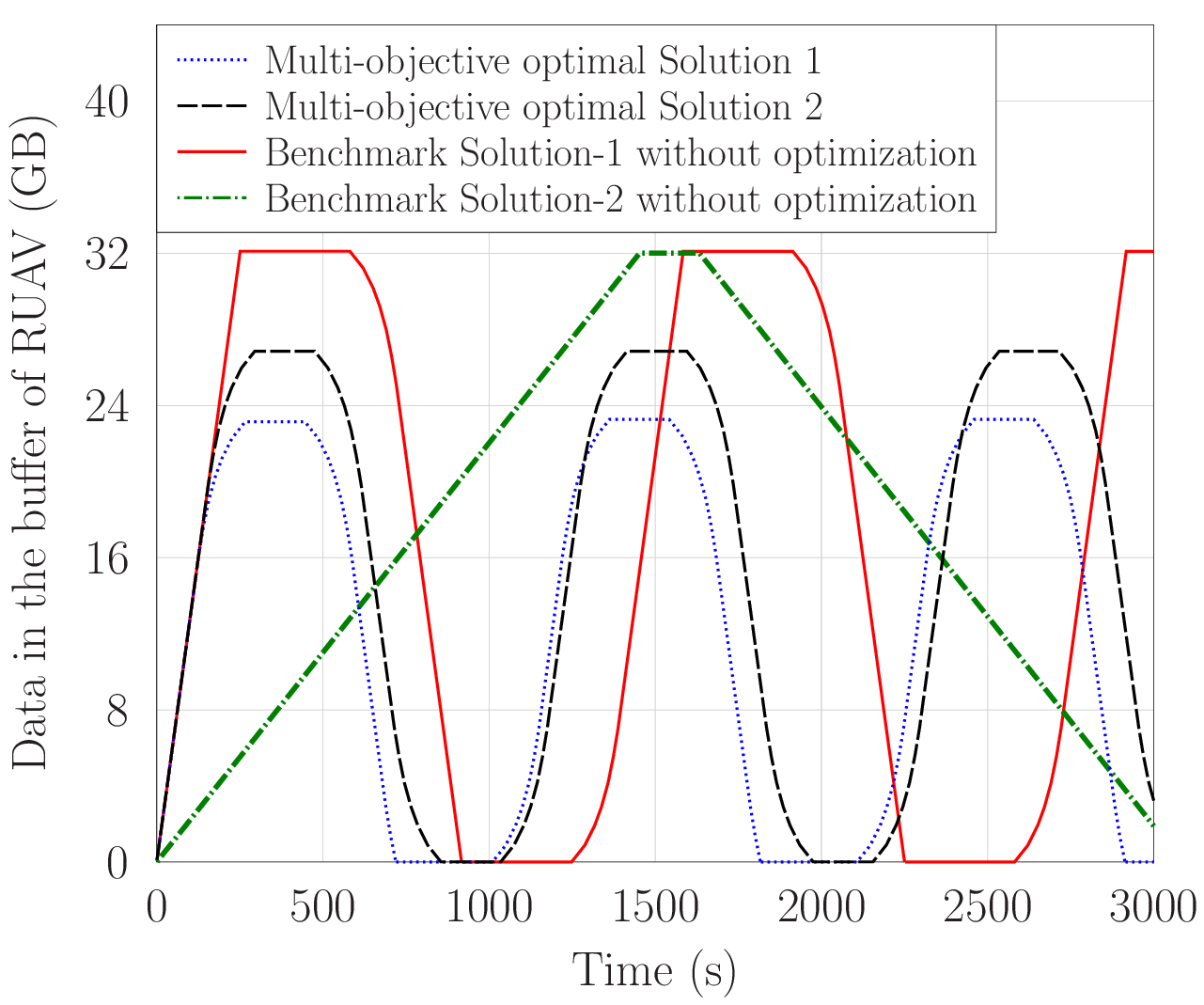} 
  \label{fig10a}
 }%
 \subfigure[Buffer size: 64G]{
  \includegraphics[width=0.47\textwidth]{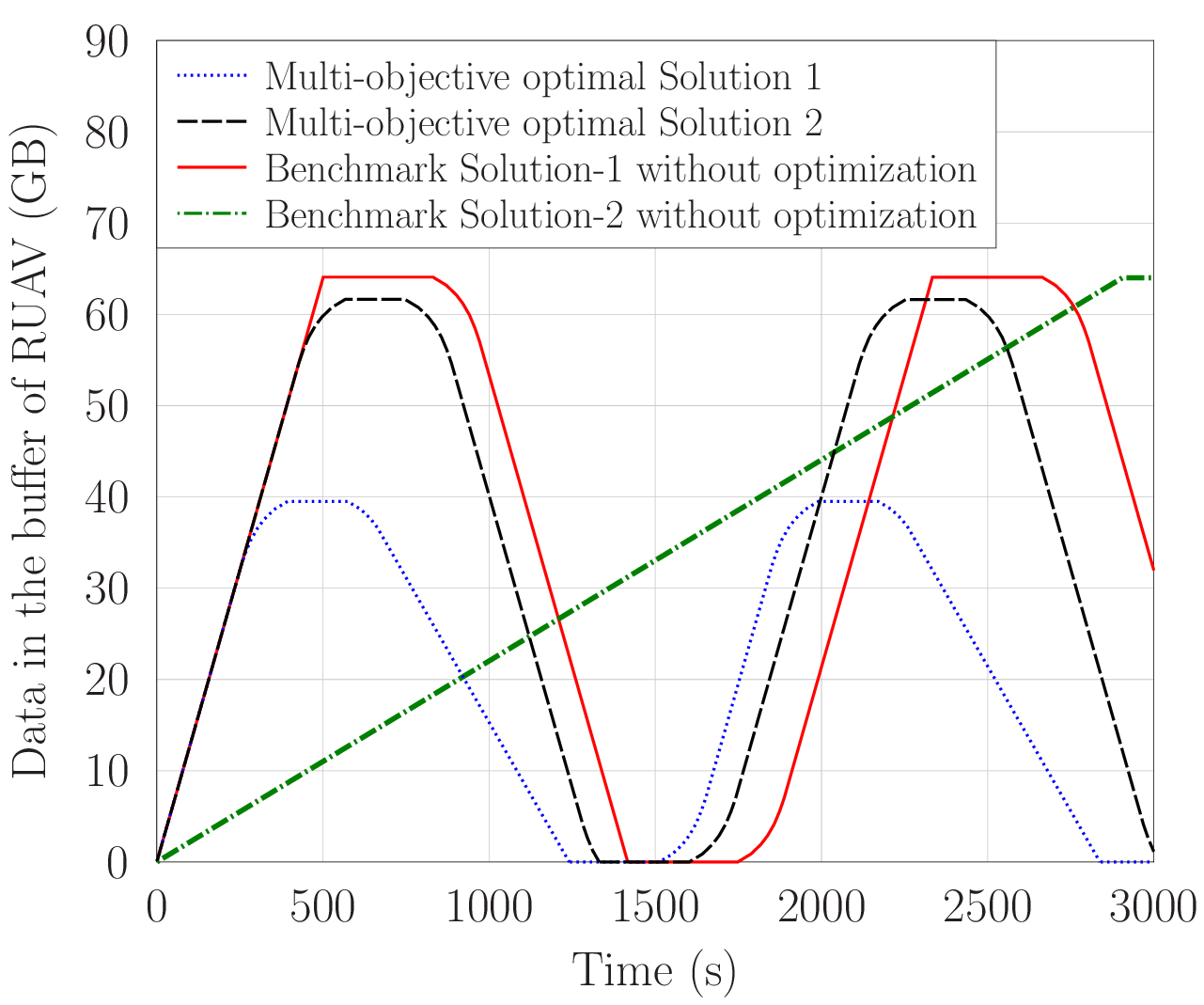}
  \label{fig10b}
 }
\end{center}
\vspace{-2mm}
\caption{Data cached in the buffer of RUAV as a function of time in \emph{Scenario~II}.}
\label{fig10}
\vspace*{-2mm}
\end{figure*}

The effective end-to-end average data rate is investigated in Fig.~\ref{fig9}, which fluctuates up and down as and when the RUAV changes its status, as illustrated in Fig.~\ref{FIG3}. Observe  from Fig.~\ref{fig9a} that although the multiple-objective Pareto optimal solutions do not always have higher effective end-to-end average data rate than the benchmark Solution 1, they reach higher effective end-to-end average data rate within 3000\,s, when the buffer size is 32\,GB. By contrast, Fig.~\ref{fig9b} shows that only the multiple-objective Pareto optimal Solution 2 reaches a higher effective end-to-end average data rate than the benchmark Solution 1 at the end of the given time period, when the buffer size is 64\,GB. As expected, when the buffer size is 64\,GB, the effective end-to-end average data rate of the benchmark Solution 2 is zero.

The amount of data cached in the buffer of the RUAV can be observed from Fig.~\ref{fig10}. It can be seen from both Fig.~\ref{fig10a} and Fig.~\ref{fig10b} that for the benchmark Solution 1, there are still lots of the data cached in the buffer of the RUAV that have not been offloaded to the GS at the end of the time period considered. Additionally, the benchmark Solution 2 has not had a chance to offload the data cached in its buffer to the GS by the end of the time period considered, when the buffer size is 64\,GB.  By contrast, both the multiple-objective Pareto optimal solutions have offloaded almost all the data to the GS at the end of the time period for both the 32\,GB and the 64\,GB buffer.

\section{Conclusions}\label{S6}

An ACM-aided and mobile relaying-assisted drone swarm network architecture, consisting of a DCDS, RUAV and GS was conceived. The DCDS is responsible for collecting data within a target area, whilst the RUAV acts as a mobile relay for hauling data from the DCDS to the GS. Furthermore, we have designed an $\epsilon$-MOGA assisted Pareto-optimization scheme associated with the four decision variables of near-DCDS loading point, near-GS offloading point, maximum factor of loading data, and minimum factor of offloading data, in order to maximize the data delivered from the DCDS to the GS, while imposing a minimum delay. We have investigated a pair of scenarios. In the first case, there are simultaneous communication links for both the DCDS-to-RUAV and the RUAV-to-GS, while for the second case, the DCDS-to-RUAV and RUAV-to-GS links do not exist concurrently. Our simulation results have demonstrated that our $\epsilon$-MOGA assisted mobile relaying is capable of delivering more data from the DSDC to the GS, while imposing minimum delay. In the scenario, when there are simultaneous DCDS-to-RUAV and the RUAV-to-GS links, our solution is capable of delivering 40.38\% and 45.38\% more data than the RUAV acting as stationary relay in the time period of 50 minutes, when the buffer sizes are 32\,GB and 64\,GB, respectively. In the scenario when the DCDS-to-RUAV and RUAV-to-GS links do not exist concurrently, our solution is capable of delivering 19.24\% and 26.86\% extra data than a non-optimized benchmark solution in the time period of 50 minutes, when the buffer sizes are 32\,GB and 64\,GB, respectively.

\end{document}